\documentclass[aps,pre,twocolumn,superscriptaddress]{revtex4-1}

\usepackage{amsmath}
\usepackage{amssymb}
\usepackage{hyperref}
\usepackage[arrow, matrix, curve]{xy}
\usepackage{bm}
\usepackage{yhmath}
\usepackage{color}
\newcommand\diff{\mathrm{d}}

\DeclareMathOperator\Imag{Im}

\begin{document}

%\preprint{}

\title{Mode-coupling theory of the glass transition for
confined fluids}



\author{Simon Lang}
\affiliation{Institut f\"ur Theoretische Physik, Universit\"at Erlangen-N\"urnberg, Staudtstra{\ss}e~7, 91058, Erlangen, Germany}

%\thanks{}

\author{Rolf Schilling}
\affiliation{Institut f\"ur Physik, Johannes Gutenberg-Universit\"at Mainz,
 Staudinger Weg 7, 55099 Mainz, Germany}

%\thanks{}

\author{Vincent Krakoviack}
\affiliation{Laboratoire de Chimie, UMR CNRS 5182, \'Ecole Normale Sup\'erieure de Lyon, 46 All\'ee d'Italie, 69364 Lyon Cedex 7, France}

\author{Thomas Franosch}
\affiliation{Institut f\"ur Theoretische Physik, Universit\"at Erlangen-N\"urnberg, Staudtstra{\ss}e~7, 91058, Erlangen, Germany}

%\thanks{}


\date{\today}

\begin{abstract}
We present a detailed derivation of a  microscopic theory for the glass transition of a liquid enclosed between two   parallel walls relying on a mode-coupling approximation. This geometry lacks translational  invariance perpendicular to the walls, which implies that the density profile and the density-density correlation function depends explicitly on the distances to the walls. We discuss the residual symmetry properties  in slab geometry and introduce a symmetry adapted complete set of  two-point correlation functions. Since the currents naturally split into components parallel and perpendicular to the  walls the  mathematical structure of the theory differs from the established mode-coupling equations in bulk. We prove that  the equations for the nonergodicity parameters still display  a covariance property  similar to  bulk liquids.
 \end{abstract}


%64.70.P- Glass transitions,
%Glass transitions, 64.70.P-
%in colloids, 64.70.pv
%in liquid crystals, 64.70.pp
%in liquids, 64.70.pm
%in metallic glasses, 64.70.pe
%in nonmetallic glasses, 64.70.ph
%in polymers, 64.70.pj
%theory and modeling of,

%Colloids, 82.70.Dd
%complex fluids, 47.57.J-
%glass transitions in, 64.70.pv

%\keywords{glass transition}

\maketitle

\section{INTRODUCTION}

Cooling or compressing a liquid usually induces  a freezing transition towards a crystal, which then corresponds to the lowest free energy state. However, in many systems the phase transformation can be circumvented  resulting in a supercooled metastable liquid where the viscosity increases by many orders of magnitude upon mild changes of temperature or density. This slowing down of transport eventually leads to the glass-transition phenomenon where structural arrest exceeds macroscopic time scales. One of the grand challenges of theoretical physics is to provide a framework that explains the microscopic mechanism and the plethora of phenomena related to the glass transition.

Significant progress in this direction has been achieved within the mode-coupling theory of the glass transition (MCT),  which was developed  by G\"otze and collaborators~\cite{Bengtzelius:1984,Goetze:Complex_Dynamics}. The theory requires only the static structure as input parameter
and then provides a  complete description of dynamic density correlations. In particular, it yields a strong slowing down of the structural relaxation upon gradual changes of the static local order, eventually leading to a structural arrest. Thus the essence of the glass transition is conceived as a dynamic breaking of ergodicity driven by the strong correlations of the constituent particles. In the vicinity of the transition MCT makes a series of non-trivial predictions that characterize the structural relaxation. The most prominent  is the emergence of two scaling laws in time, a phenomenon
that does not appear to have an analog in other fields of physics.
The first scaling law describes   the dynamics close to a plateau value, also referred to as nonergodicity parameter or glass form factor, and a factorization property of the space and time dependence  is predicted. The decay from the plateau to zero obeys a second scaling law (time-temperature superposition principle) characterized by  stretched relaxation functions.

Numerous aspects of MCT  have been tested successfully~\cite{Goetze:1999};  examples include depolarized-light scattering, which nicely displays the enhancement of a minimum in the first
 scaling regime~\cite{Du:1994,Franosch:1997b,Singh:1998},  colloidal glass-forming systems~\cite{Megen:1993,Megen:1998} exhibiting  the wave-number-dependent structural relaxation, and computer simulations on binary mixtures ~\cite{Kob:1994,Kob:1995} revealing scaling behavior in the vicinity of the plateau.
%Despite of the success of the theory for simple one-component systems or mixtures, extension of the mode-coupling approach~\cite{Franosch:1998} to more complex systems have been rare until recently.

The success of the theory for simple one-component systems or mixtures is encouraging to advance the mode-coupling approach of the glass transition also to more complex situations, adding new degrees of freedom, varying the dimension or introducing confinement.  The mode-coupling theory  has been applied successfully   to two dimensions~\cite{Bayer:2007,Hajnal:2009,Hajnal:2011}. Similarly, the properties of  MCT in arbitrarily high dimensions have been discussed~\cite{Schmid:2010, Miyazaki:2010, Schmid:2011, Miyazaki:2011} to infer if the theory becomes of mean-field type in a well-defined manner. Whereas these works describe simple liquids composed of structureless particles,
already rigid linear~\cite{Scheidsteger:1997,Franosch:1997,Scheidsteger:1998,Kaemmerer:1998,Kaemmerer:1998a} and arbitrarily shaped ~\cite{Fabbian:1999}  molecules require the use of symmetry-adapted tensor fluctuation densities to account for the orientational degrees of freedom. Then, the intermediate scattering function generalizes to a matrix-valued correlation  function accompanied by a splitting of the currents,
which introduces subtle new mathematical properties~\cite{Schilling:2002}.

A challenge for the theory is introduced by exposing the glass-forming liquid to complex geometries or external potentials~\cite{Ramaswamy:2011} and walls. A mode-coupling theory combining aspects of quenched disorder and interaction induced vitrification has been elaborated ~\cite{Krakoviack:2005,Krakoviack:2007,Krakoviack:2009,Krakoviack:2011}, which predicts an intriguing interplay of the glass transition driven by the strong mutual interactions of the fluid particles
and the localization transition  induced by the obstruction by the frozen matrix.
However, there a divergent length scale with long-wavelength anomalies~\cite{Krakoviack:2009,Schnyder:2011} emerges and a  refined description requires  concepts from
critical phenomena as has been worked out for the  Lorentz problem~\cite{Lorentz_PRL:2006,Lorentz_JCP:2008,Franosch:2011}.

 The response of the time-dependent density correlation function to small local perturbations requires one to consider inhomogeneous mode-coupling equations~\cite{Biroli:2006}, and the emergence of  a hidden divergent length scale has been predicted.

A great deal of experimental work and computer simulations has been devoted to confine the liquid~\cite{Workshop_Confinement:2000} to a narrow slab to investigate the role of cooperativity and dynamic heterogeneities for the slowing down of transport processes. The interaction of the liquid with the walls has  a crucial influence on the glass transition, e.g., for rough walls an increase of the critical temperature in comparison to bulk liquids has been reported~\cite{Scheidler:2000,Scheidler:2000a,Scheidler:2004, Teboul:2002}. For smooth repulsive walls an increase, e.g., for soft sphere mixtures~\cite{Fehr:1995}, was found as well, whereas the opposite  was identified for polymers~\cite{Varnik:2000,Varnik:2002,Baschnagel:2005}. Experimental results using confocal microscopy on colloidal hard-sphere suspensions between two smooth walls reveal a decrease of the critical packing fraction  and a slower dynamics at the walls in lateral direction~\cite{Nugent:2007,Eral:2009,Eral:2011}. Recently, an additional slowing down of motion has been reported due to an enhancement of effective surface roughness  by immobilized particles~\cite{Edmond:2010}.
Computer simulations for the diffusive dynamics of  hard spheres   found that the diffusivity displays peculiar behavior as the distance of the plates is varied~\cite{Mittal:2008}, oscillating similarly to static quantities like the excess entropy obtained by density-functional theory~\cite{Goel:2009}. Molecular dynamics simulations for  water confined to silica pores have been analyzed in terms of the universal aspects of MCT for bulk liquids close to the glass transition~\cite{Gallo:2000a,Gallo:2000b,Gallo:2012}.

Recently, we have introduced a mode-coupling theory for the glass transition in slab geometry based on symmetry adapted density fluctuation modes. For the case of a hard-sphere liquid
a nontrivial interplay between the length scale of confinement and the average distance between the particles has been predicted~\cite{Lang:2010}. Thereby a facilitation of the glass transition
 close to half-integer values of the distance with respect to the hard-sphere diameter was found in the theory. In contrast, at distances near  integer multiples of the particle diameter
 the liquid phase remains favored for higher packing fractions,
which allows us to be interpreted as a manifestation of commensurability effects.

In this article we  provide a detailed derivation of the mode-coupling theory for confined fluids. The theory describes the dynamics of simple fluids confined by two parallel flat hard  walls. We discuss the residual symmetries and  design  a symmetry-adapted complete set of Fourier modes to
decompose the density fluctuations in real space. Then the density-density correlation function is expanded in terms of a matrix-valued intermediate scattering function and exact equations of motion are derived using the Zwanzig-Mori projection-operator formalism. The currents naturally split into  components parallel and perpendicular to the surfaces.  The mode-coupling approximation is applied for the force kernel  leading to a set of closed equations of motion for the generalized intermediate scattering function. The theory requires as input the density profiles and the static structure factors of the confined liquid.

The MCT equations of the glass transition for simple one-component liquids are covariant under a linear transformation of the time-dependent density correlator~\cite{Goetze:1995,Goetze:Complex_Dynamics}. This covariance, which has been proven to be valid for multicomponent systems as well~\cite{Franosch:2002}, has strong implications for the properties  of the solutions of the MCT equations~\cite{Goetze:1995,Goetze:Complex_Dynamics,Franosch:2002}. The decomposition of the current density for confined liquids into a parallel and a perpendicular component leads to MCT equations of a different mathematical structure than for one- or multicomponent liquids. Here we provide a first step to demonstrate that some of these properties hold also within the mode-coupling theory for confined liquids by showing the covariance property of the MCT equations for the nonergodicity parameters. In particular, we prove the existence of one of its solutions distinguished by a maximum principle.

\section{MODEL AND INVARIANCE PROPERTIES}

We consider a simple liquid comprised of $N$ identical particles of mass $m$ without inner degrees of freedom enclosed between two flat, hard, and parallel walls, which are separated by a distance $L$. The area of the wall surfaces is denoted by $A$, and a thermodynamic limit $A\to \infty, N\to \infty$  is anticipated, such that the area density $n_0:=N/A$ and the wall separation $L$ remain
constant.
Adapted to the geometrical constraints  a coordinate system is introduced such   that the $z$ axis is perpendicular to  the hard surfaces located at $z=\pm L/2$.
Here we adopt the convention that the location of the surfaces are taken to confine  the centers of the particles  to $|z| \leq L/2$. In a real experiment the particles cannot approach the  plates further than a hard core radius $\sigma/2$, e.g., for hard spheres the effective distance of plates is then $H = L+\sigma$. For the development of the theory it is convenient to use $L$ as the relevant confinement length.

 In the following the  in-plane coordinates are abbreviated by $\vec{r}=(x,y)$. The positions of the centers  and momenta of the $N$-particle system are  specified by $\{ \vec{x}_{n}\} = \{(\vec{r}_{n},z_{n})\} = (\vec{x}_{1},...,\vec{x}_{N})$ and   $\{ \vec{p}_{n}\} = \{(\vec{P}_{n},P_{n}^{z})\} = (\vec{p}_{1},...,\vec{p}_{N})$  with  the in-plane momenta $\{ \vec{P}_{n}\}$. The positions and momenta of the particles evolve according to Newton's equations of motion and  the corresponding  Hamilton function is given by
\begin{equation}
 H(\{\vec{x}_{n}\},\{\vec{p}_{n}\})=\sum_{n=1}^{N} \frac{\vec{p}_{n}^2}{2m}+V(\{\vec{x}_{n}\})+U(\{z_{n}\}).
\end{equation}
For simplicity the mutual interaction between the particles is assumed to be  pairwise additive
\begin{equation}
V(\{\vec{x}_{n}\})=\sum_{n<m}^{N} \mathcal{V}(|\vec{x}_{n}-\vec{x}_{m}|),
\end{equation}
such that the two-particle interaction preserves linear momentum and angular momentum.
The walls confine the  particles between the flat surfaces. Additionally, a specific particle-wall interaction such as adsorption induced by hydrophilic or adhesive, respectively, hydrophobic or cohesive forces, can be included,
\begin{equation}
U(\{z_{n}\})=\sum_{n=1}^{N}\mathcal{U}(z_{n}),
\end{equation}
where
\begin{equation}
\mathcal{U}(z)=
\begin{cases}
    \mathcal{U}^{W}(z) & \text{for } |z| \leq L/2,  \\
  \infty & \text{for } |z| > L/2.
\end{cases}
\end{equation}
For the case of identical walls, the specific interaction displays the additional symmetry $\mathcal{U}^W(-z) = \mathcal{U}^W(z)$.

Note that the wall constraint drastically changes the structure and dynamics of the liquid, which cannot be treated within perturbation theory. Therefore, a linear response approach as suggested in~\cite{Biroli:2006} is not suitable to capture the induced changes.

The physical quantities characterizing the dynamics of the confined liquid reflect the symmetries of the equilibrium distribution  and the corresponding time evolution.  Whereas a  bulk system is on average isotropic and  translationally invariant, and displays space- and time-inversion symmetry, the walls reduce the number of symmetry transformations.
These residual symmetries are determined by investigating the invariance properties of the Hamilton function. A phase-space transformation ${\cal O}$ is called a symmetry of the  system if it leaves the Hamilton function   invariant $H(\{\mathcal{O} \vec{x}_{n}\},\{\mathcal{O} \vec{p}_{n}\})=H(\{\vec{x}_{n}\},\{\vec{p}_{n}\})$, at least in the  limit of large system sizes.  The set  of all
space-time symmetries compatible with the constraint is generated by the elementary transformations
\begin{align}
 \mathbb{O}=&
\{\mathcal{T}_{\vec{d}},\mathcal{R}_{z}(\alpha), \mathcal{P}_{xz}, \mathcal{P}_{yz}, \{ \Pi \}, \mathcal{T}_{t}, \mathcal{T}_{\pm} \},
\end{align}
where  ${\mathcal{T}_{\vec{d}}}$ are arbitrary in-plane translations by a vector  $\vec{d}=(d_x,d_y,0)$, $\mathcal{R}_{z}(\alpha)$ are rotations around the $z$ axis  by an angle $\alpha\in[0,2\pi)$. The
elements $\mathcal{P}_{xz}$ and  $\mathcal{P}_{yz}$ correspond to reflections at the $x$-$z$ and the $y$-$z$ plane, respectively. A permutation $\Pi$ of the particle labels $1,2, \dots,N$ also leaves the statistical properties as well as the dynamics
unchanged.  Furthermore, shifts ${\mathcal T}_t$ by a time $t$ or time reversal  $\mathcal{T}_\pm$ leave  the dynamical laws invariant, thus
do not change correlation functions.

For identical  walls the
symmetry group is larger and is generated by
\begin{align}
 \mathbb{O}_\text{sym}=&  \mathbb{O} \cup
\{\mathcal{R}_{\vec{n}}(\pi), \mathcal{I}, \mathcal{P}_{xy}\},
\end{align}
where $\mathcal{R}_{\vec{n}}(\pi)$ are rotations by an angle $\pi$ around an axis $\vec{n}=(n_x,n_y,0)$ in the $x$-$y$ plane. A space inversion is generated by $\mathcal{I}$, and $\mathcal{P}_{xy}$ indicates
a reflection at the dividing $x$-$y$ plane.

\section{CORRELATION FUNCTIONS IN REAL AND WAVE-NUMBER SPACE}

In this section we present the correlation functions in real space adapted to the confined geometry.
We start with the most basic quantity describing a liquid, the microscopic particle density
\begin{equation}\label{eq:micrho}
 \rho(\vec {r},z,t)= \sum _{n=1}^{N} \delta [\vec{r}-\vec{r}_{n}(t)] \delta [z-z_{n}(t)],
\end{equation}
where the dependence on the initial position in phase space  is omitted for simplicity. Due to translational symmetry parallel to the surfaces,  ${\mathcal{T}_{\vec{d}}}$, the equilibrium density varies only in the $z$-direction
\begin{equation}
 n(z)=\langle \rho(\vec {r},z,t)\rangle .
\end{equation}
Here the angle brackets $\langle \cdot \rangle$ indicate canonical averaging over the initial conditions in phase space.
We then introduce the fluctuations $\delta\rho(\vec {r},z,t) :=\rho(\vec {r},z,t)-n(z)$ and define the  density-density correlation function, which corresponds to the Van Hove  function~\cite{Hansen:Theory_of_Simple_Liquids},
\begin{equation}
 G(|\vec{r}-\vec{r}'|,z,z',t):=\frac{1}{n_{0}} \langle \delta\rho(\vec{r},z,t)\delta\rho(\vec{r}',z',0)\rangle.
\end{equation}
By translational and rotational symmetry in plane, $\mathcal{T}_{\vec{d}}$ and $\mathcal{R}_{z}(\alpha)$, the Van Hove function only depends on the modulus of the in-plane distance $|\vec{r}-\vec{r}'|$. Furthermore, by time reversal symmetry, ${\mathcal T}_\pm$, it is an even function of time
\begin{equation}
G(|\vec{r}-\vec{r}'|,z,z',t) =G(|\vec{r}-\vec{r}'|,z,z',-t),
\end{equation}
and by time translational symmetry, ${\mathcal T}_t$, it is symmetric with respect to interchanging the positions
\begin{equation}
G(|\vec{r}-\vec{r}'|,z,z',t) =G(|\vec{r}'-\vec{r}|,z',z,t).
\end{equation}
For the case of identical walls, the density profile is symmetric, $n(z) = n(-z)$, and the Van Hove function is invariant under simultaneous reflection of $z$ and $z'$,
\begin{equation}
   G(|\vec{r}-\vec{r}^{\prime}|,z,z^{\prime},t) =G(|\vec{r}-\vec{r}^{\prime}|,-z,-z^{\prime},t).
\end{equation}

We expand all quantities in terms of  symmetry-adapted  Fourier modes. For the $z$ direction we employ a discrete set of Fourier modes $\exp(-\text{i} Q_\mu z)$ with wave numbers $Q_{\mu}=2\pi \mu/L, \mu \in \mathbb{Z}$. These constitute a complete set $\sum_\mu \exp(\text{i} Q_\mu z) \exp(-\text{i} Q_\mu z') = L \delta(z-z')$
of orthogonal functions in the finite interval $[-L/2,L/2]$: $\int_{-L/2}^{L/2}  \exp(\text{i} Q_\mu z) \exp(-\text{i} Q_\nu z) \diff z= L \delta_{\mu\nu}$.
Hence the equilibrium density profile is expanded in discrete modes
\begin{equation}
 n(z) = \frac{1}{L}\sum_\mu n_\mu \exp(-\text{i} Q_\mu z),
\end{equation}
where sums over greek subscript indices are to be taken over all integer numbers $\mathbb{Z}$.  The  corresponding Fourier coefficients are obtained as
\begin{equation}\label{eq:Fourierdensity}
 n_{\mu}= \int\limits_{-L/2}^{L/2} \diff z \ n(z) \exp(\text{i}  Q_{\mu}z).
\end{equation}
Since $n(z)$ is real, $n_\mu = n_{-\mu}^*$.
We shall also need  the local specific volume $v(z):=1/n(z)$. By the convolution theorem, the Fourier coefficients fulfill
\begin{equation}\label{eq:sum}
\sum_\kappa n_{\mu-\kappa} v_{\kappa-\nu} = \sum_\kappa n_{\mu-\kappa}^* v_{\kappa-\nu}^* =  L^2 \delta_{\mu\nu}.
\end{equation}
For symmetric walls the coefficients are  real,
\begin{equation}
n_{\mu}= n^{*}_{\mu}, \qquad v_{\mu}=v^{*}_{\mu},
\end{equation}
and by the previous relation, they are also symmetric, $n_\mu = n_{-\mu}, v_\mu = v_{-\mu}$.

We decompose  the spatial dependence  parallel to the surfaces into ordinary plane waves, $\text{e}^{-\text{i} \vec{q} \cdot \vec{r}}$, where the wave vectors $\vec{q} = (q_x,q_y)$ are treated
initially as discrete $ (q_x,q_y) \in (2\pi/\sqrt{A}) \mathbb{Z}^2$. For example,  the microscopic density is
\begin{equation}
 \rho(\vec{r},z,t) = \frac{1}{A} \sum_{\vec{q}} \frac{1}{L} \sum_\mu \rho_\mu(\vec{q},t) \exp(-\text{i} Q_\mu z) \text{e}^{-\text{i} \vec{q} \cdot \vec{r}}.
\end{equation}
If one performs the thermodynamic limit, such that $\vec{q}$ becomes a continuous variable, sums are replaced by integrals $(1/A) \sum_{\vec{q}} \ldots \to  (2\pi)^{-2}\int \diff^2 \vec{q}  ...$  as usual. The fundamental quantities of interest are the expansion coefficients $\rho_\mu(\vec{q},t)$ called density modes,
\begin{equation}
 \rho_\mu (\vec{q},t)=\sum\limits _{n=1}^{N}\exp [ \text{i} Q_\mu z_n (t)] \, \text{e}^{\text{i} \vec{q} \cdot \vec{r}_n(t)},
\end{equation}
with corresponding fluctuations
$ \delta \rho_\mu (\vec{q},t)=\rho_\mu (\vec{q},t)-\langle\rho_\mu (\vec{q},t)\rangle$. Since $\langle\rho_\mu (\vec{q},t)\rangle  = A n_{\mu}\delta_{{\vec{q},0}}$\label{Mittelueberrho} the correction is relevant only for vanishing wave vector parallel to the confinement.

Expressing the density fluctuations in real space $\rho(\vec{r},z,t)$ by its Fourier decomposition yields an expansion of the corresponding Van Hove  function
\begin{align}
 G(|\vec{r}-\vec{r}'|,z,z',t) =& \frac{1}{A} \sum_{\vec{q}} \frac{1}{L^2} \sum_{\mu \nu}  S_{\mu \nu}(\vec{q},t)  \nonumber \\
& \times \exp\left[ \text{i}(Q_\mu z- Q_\nu z^{\prime}) \right] \text{e}^{\text{i} \vec{q} \cdot (\vec{r}-\vec{r}^{\prime})} ,
\end{align}
in terms of an infinite matrix  $[{\mathbf S}(q,t)]_{\mu\nu}=S_{\mu
\nu}(q,t)$, which generalizes the intermediate scattering function,
\begin{equation}\label{eq:density-density-correlation}
 S_{\mu\nu}(q,t)=\frac{1}{N} \langle \delta \rho_\mu (\vec{q},t)^{*}\delta \rho_\nu (\vec{q},0)\rangle.
\end{equation}
The translational invariance along the direction of the walls manifests itself in the appearance of  a single wave vector $\vec{q}$, whereas perpendicularly two indices are required. Furthermore it depends only on the magnitude $q=|\vec{q}|$ due to rotational invariance in the plane, $\mathcal{R}_z(\alpha)$.

Reversely, the generalized intermediate scattering function is obtained from the van  Hove function by Fourier transform,
\begin{align} \label{eq:Fourier}
 S_{\mu\nu}(q,t)=&
 \int\limits_{-L/2}^{L/2}\!\!\!\!\diff z \!\! \int\limits_{-L/2} ^{L/2}\!\!\!\!\diff z^{\prime}\int\limits_{A} \! \diff (\vec{r}-\vec{r}') G({|\vec{r}-\vec{r}^{\prime}|,z,z^{\prime},t})\nonumber  \\
 & \times \exp\left[- \text{i}(Q_\mu z- Q_\nu z') \right] \text{e}^{-\text{i} \vec{q} \cdot (\vec{r}-\vec{r}^{\prime})}.
\end{align}
The initial value $S_{\mu\nu}(q) :=S_{\mu\nu}(q,t=0)$ characterizes the equilibrium structure of the fluid in the slit and will be referred to as generalized static structure factor.  Note, that the hermitian matrix $\mathbf{S}(q)\succeq 0$ is non-negative, i.e., for any set of complex numbers $y_\nu$, the inequality $\sum_{\mu \nu} y_\mu^* S_{\mu\nu}(q) y_\nu \geq 0$ holds.

The space and time symmetries imply  relations between the matrix elements of $S_{\mu\nu}(q,t)$. The translational $\mathcal{T}_{\vec{d}}$ and rotational symmetry $\mathcal{R}_{z}(\alpha)$ have already been exploited.
By time  reversal symmetry $\mathcal{T}_\pm$ and time
translation $\mathcal{T}_t$, the intermediate scattering function is even in time and the matrix is hermitian,
\begin{equation}
 S_{\mu\nu}(q,t) =S_{\mu\nu}(q,-t) = S_{\nu\mu}(q,t)^*.
\end{equation}
For symmetric walls, the  inversion symmetry $\mathcal{I}$ yields
\begin{equation}
 S_{\mu\nu}(q,t) =S_{\mu\nu}(q,t)^* = S_{-\mu-\nu}(q,t),
\end{equation}
i.e., the matrices are real symmetric and invariant under simultaneous change of sign of the mode indices.

The conservation of the particle number  within the slit geometry is encoded in  the continuity  equation
\begin{equation}\label{eq:field}
\frac{\partial}{\partial t} \rho(\vec{r},z,t)+ \vec{\nabla}\cdot\vec{j}(\vec{r},z,t)=0,
\end{equation}
where the microscopic particle current density  is given by
\begin{equation}
\vec{j}(\vec{r},z,t)=\sum _{n=1}^{N} \frac{\vec{p}_{n}(t)}{m} \delta [\vec{r}-\vec{r}_{n}(t)] \delta [z-z_{n}(t)].
\end{equation}
By the symmetry of the system the currents split naturally into
current densities parallel,
\begin {equation}
\vec{j}^{\parallel}(\vec{r},z,t)=\sum _{n=1}^{N} \frac{\vec{P}_{n}(t)}{m} \delta [\vec{r}-\vec{r}_{n}(t)] \delta [z-z_{n}(t)],
\end {equation}
and perpendicular to the surfaces,
\begin {equation}
j^{\perp}(\vec{r},z,t)=\sum _{n=1}^{N} \frac{P^{z}_{n}(t)}{m} \delta [\vec{r}-\vec{r}_{n}(t)] \delta [z-z_{n}(t)].
\end {equation}
In this work we need only the longitudinal components that contribute to the particle conservation law. Then  the divergence  $\vec{\nabla}\cdot \vec{j}(\vec{r},z,t)=\vec{\nabla}_{\vec{r}}\cdot~\vec{j}^{\parallel}(\vec{r},z,t)+\nabla_{z} j^{\perp}(\vec{r},z,t)$ consists of two decay channels, with currents that are represented in the Fourier domain,
\begin{equation}\label{eq:currents}
 j_{\mu}^{\alpha}(\vec{q},t)\!=\! \frac{1}{m} \!\sum\limits_{n=1}^{N}b^{\alpha}(\hat{\vec{q}} \cdot {\vec{P}}_{n}(t),P_{n}^{z}(t)) \exp[i Q_\mu z_n (t)] \, \text{e}^{i \vec{q} \cdot \vec{r}_n(t)}.
\end{equation}
Here we  abbreviate the unit vector $\hat{\vec{q}} = \vec{q}/q$ and introduce the selector $b^\alpha(x,z) = x \delta_{\alpha,\parallel} + z\delta_{\alpha,\perp}$, which will simplify the subsequent manipulations. Consequently, a spatial Fourier expansion of Eq.~\eqref{eq:field} leads to   the continuity equation for the  density and current density modes
 \begin{equation}\label{eq:Continuity}
  \partial_t \rho_\mu (\vec{q},t)= \text{i} \sum_{\alpha=\parallel,\perp} b^{\alpha}(q,Q_{\mu}) j_{\mu}^{\alpha}(\vec{q},t).
\end{equation}
It is instructive to consider also  the current density correlator matrix $[\bm{\mathcal{J}}(q,t)]^{\alpha\beta}_{\mu\nu}=\mathcal{J}_{\mu \nu}^{\alpha \beta}(q,t)$,  with matrix elements defined by
\begin{equation}\label{eq:currentcorrelator}
 \mathcal{J}_{\mu \nu}^{\alpha \beta}(q,t)= \frac{1}{N} \langle  j_{\mu}^{\alpha }(\vec{q},t)^* j_{\nu}^{\beta}(\vec{q},0)\rangle.
\end{equation}
In particular, its initial value  $\mathcal{J}^{\alpha\beta}_{\mu\nu}(q)=\mathcal{J}^{\alpha\beta}_{\mu\nu}(q,t=0)$ can be evaluated explicitly (see Appendix \ref{sec:Current})
\begin{equation}\label{eq:currents_static}
 \mathcal{J}^{\alpha\beta}_{\mu\nu}(q)=\frac{k_{B}T}{m}\frac {n_{\mu-\nu}^*}{n_{0}}\delta_{\alpha\beta}.
\end{equation}
By Eq. (\ref{eq:sum}) its inverse matrix can be expressed in terms of the local specific volume
\begin{equation}\label{eq:currents_static_inverse}
 [\bm{\mathcal{J}}^{-1}(q)]^{\alpha\beta}_{\mu\nu}=\frac{m}{k_{B}T} n_{0} \frac{v_{\mu-\nu}^*}{L^{2}}\delta_{\alpha\beta}.
\end{equation}
Applying the same reasoning as above, one easily derives the symmetry relations for the current-current correlator,
\begin{equation}
\mathcal{J}^{\alpha\beta}_{\mu\nu}(q,t) = \mathcal{J}^{\alpha\beta}_{\mu\nu}(q,-t) = \mathcal{J}^{\beta\alpha}_{\nu\mu}(q,t)^*  ,
\end{equation}
and for symmetric walls additionally,
\begin{equation}
 \mathcal{J}^{\alpha\beta}_{\mu\nu}(q,t) = \mathcal{J}^{\alpha\beta}_{\mu\nu}(q,t)^* = \mathcal{J}^{\beta\alpha}_{-\nu-\mu}(q,t).
\end{equation}
The emergence of the channel indices ($\alpha,\beta$) for the current density correlator matrix, that represent the splitting of the currents into a parallel and perpendicular component, occurs in the same spirit as has been introduced for molecular liquids ~\cite{Scheidsteger:1997} or a single molecular solute~\cite{Franosch:1997}. There, the currents naturally split into a translational and an orientational part.

\section{ZWANZIG-MORI PROJECTION-OPERATOR FORMALISM}
In this section the equations of motion for the generalized intermediate scattering function  are  derived with the help of the Zwanzig-Mori projection-operator formalism~\cite{Forster:Hydrodynamic_Fluctuations,Goetze:Complex_Dynamics}. 

The dynamics is driven by Newton's equations of motion, which implies that the time evolution of  phase-space functions $A(t) \equiv A(\{\vec{p}_{n}(t)\}, \{\vec{x}_{n}(t)\})$ is obtained by $\partial_t A(t) = \{ A(t), H \} \equiv \text{i} {\cal L} A(t)$, where ${\cal L}$  is referred to as the  Liouville operator. The formal solution then reads $A(t)  = \exp(\text{i} {\cal L} t) A$, where we adopt the convention that if no argument is provided the phase space function refers to the initial time $t=0$.

The set of fluctuating phase space functions is naturally equipped with a Hilbert space structure via the Kubo scalar product $\langle  A |  B \rangle \equiv \langle \delta A^{*} \delta B\rangle$ as correlation functions between fluctuations  $\delta A = A - \langle A \rangle$. One easily convinces oneself that the Liouville operator is hermitian with respect to the Kubo scalar product, see, e.g., ~\cite{Goetze:Complex_Dynamics} for the mathematical rigor. Dynamic correlation functions
can then be represented as matrix elements  $\langle \delta A(t)^* \delta B \rangle = \langle A | {\cal R}(t) | B \rangle$ of the backwards-time evolution operator ${\cal R}(t) = \exp(- \text{i} {\cal L} t)$.  The projection-operator formalism relies on an exact reformulation
of the operator identity $\partial_t \mathcal{R}(t) = -\text{i} \mathcal{L} \mathcal{R}(t)$ to
\begin{align}\label{eq:operator_identity}
&\partial_{t} \mathcal{P}\mathcal{R}(t)\mathcal{P}+\text{i}\mathcal{P}\mathcal{L}\mathcal{P}\mathcal{R}(t)\mathcal{P}\nonumber \\
&+\int_{0}^{t} \diff t'\mathcal{P}\mathcal{L}\mathcal{Q}  \exp[- \text{i}{\cal Q} {\cal L}\mathcal{Q} (t-t')] {\cal Q } {\cal L}\mathcal{P}\mathcal{R}(t')\mathcal{P}=0,
\end{align}
valid for any orthogonal projection operator ${\cal P}$; see  Appendix \ref{sec:operator-identity}. Here   $ {\cal Q}= \mathbf{1} -{\cal P}$
denotes the projection onto the orthogonal complement, and ${\cal R}_{\cal Q}(t) =  \exp(- \text{i} {\cal Q} {\cal L} {\cal Q} t)$ is referred to as the reduced backwards-time evolution operator.

Here we derive a formally exact equation of motion for the generalized intermediate scattering function $S_{\mu\nu}(q,t) = \langle  \rho_\mu(\vec{q}) | {\cal R}(t) | \rho_\nu(\vec{q}) \rangle/N$.
First we use the density  as distinguished variable and introduce the projector
\begin{equation}
 \mathcal {P}_{\rho}=\frac{1}{N}\sum_{\vec{q}}\sum_{\mu\nu}|\rho_{\mu}(\vec{q})\rangle [\mathbf{S}^{-1}(q)]_{\mu\nu} \langle  \rho_{\nu}(\vec{q})|,
\end{equation}
with corresponding orthogonal projection operator $\mathcal{Q}_{\rho}=\mathbf{1}  -\mathcal{P}_{\rho}$. Sandwiching the operator identity Eq.~\eqref{eq:operator_identity} between the distinguished variables, one derives the first equation of motion
\begin{equation}\label{eq:eom1}
 \dot{S}_{\mu\nu}(q,t) + \sum_{\kappa\lambda}\!\! \int_0^t  \! \! K_{\mu \kappa}(q,t-t') [\mathbf{S}^{-1}(q)]_{\kappa\lambda} S_{\lambda\nu}(q,t')\diff t' = 0.
\end{equation}
Here we observed that in Newtonian dynamics $\langle \rho_\mu(\vec{q}) | {\cal L} | \rho_\nu(\vec{q} ) \rangle =0 $, such that the second term in Eq.~\eqref{eq:operator_identity} does not contribute.
The third term can be simplified using ${\cal Q}_\rho {\cal L} | \rho_\mu(\vec{q}) \rangle ={\cal L} | \rho_\mu(\vec{q}) \rangle$ and leads to
 the memory kernel $\mathbf{K}(q,t)$ with matrix elements,
\begin{equation}\label{eq:K}
 K_{\mu\nu}(q,t)=\frac{1}{N} \langle \mathcal{L}\rho_{\mu}(\vec{q}) | \mathcal{R}_{\mathcal{Q}_{\rho}}(t)  |
\mathcal{L} \rho_{\nu}(\vec{q})\rangle.
\end{equation}
In contrast to bulk systems, the current densities display two relaxation channels, one in plane and one in the perpendicular direction.
 By  the particle conservation law, Eq.~\eqref{eq:Continuity}, the memory kernel  naturally splits into four parts,
\begin{equation}\label{eq:split}
 K_{\mu\nu}(q,t)=\sum_{\alpha\beta=\parallel,\perp}b^{\alpha}(q,Q_{\mu}) \mathcal{K}^{\alpha\beta}_{\mu\nu}(q,t)
 b^{\beta}(q,Q_{\nu}),
\end{equation}
with the reduced current-current correlation matrix
\begin{equation}
 \mathcal{K}^{\alpha\beta}_{\mu\nu}(q,t)=\frac{1}{N}\langle j_{\mu}^{\alpha}(\vec{q})|\mathcal{R}_{\mathcal{Q}_{\rho}}(t)|j_{\nu}^{\beta}(\vec{q})\rangle.
\end{equation}
The symmetries of $\mathcal{K}^{\alpha \beta}_{\mu\nu}(q,t)$ are identical to the current-current correlation function ${\cal J}^{\alpha\beta}_{\mu\nu}(q,t)$. Its initial value coincides with the equilibrium static current-current correlator, Eq.~\eqref{eq:currents_static}. Solving the first equation of motion, Eq.~\eqref{eq:eom1}, for $S_{\mu\nu}(q,t)$ to second order in the lag time $t$, one derives the
short-time expansion
\begin{equation}
 S_{\mu\nu}(q,t)=S_{\mu \nu}(q)-\frac{1}{2}  \frac{k_{B}T}{m} \frac{n^{*}_{\mu-\nu}}{n_{0}}(q^2+Q_{\mu} Q_{\nu}) t^2 + \mathcal{O}(t^4).
\end{equation}
The parallel relaxation gives rise to a term for the motion along the plates and a second one for the flow perpendicular to the confinement. Note that different mode indices contribute in a nontrivial way to the decay of $S_{\mu\nu}(q,t)$ already at order ${\cal O}(t^2)$, due to the breaking of translational symmetry.

The reduced current correlator is not suited as a starting point for approximations for the slow dynamics. Rather, we employ a second Zwanzig-Mori step for the reduced backwards-time evolution operator ${\cal R}_{{\cal Q}_\rho}(t)$, hence we make the replacement ${\cal L} \mapsto {\cal Q}_\rho {\cal L} {\cal Q}_\rho$ in the operator identity Eq.~\eqref{eq:operator_identity}. The new orthogonal projector is constructed from the current kets $|j_\mu^\alpha(q) \rangle$. Since currents to different relaxation channels are mutually orthogonal, Eq.~\eqref{eq:currents_static}, the projector splits into  two commuting orthogonal components,
\begin{equation}
 \mathcal{P}_{j}=\sum_{\alpha=\parallel,\perp}\mathcal{P}_{j}^{\alpha},
\end{equation}
where the individual projections are represented by
\begin{equation}
 \mathcal{P}_{j}^{\alpha}=\frac{1}{N}\sum_{\vec{q}}\sum_{\mu\nu}|j_{\mu}^{\alpha}(\vec{q})\rangle [\bm{\mathcal{J}}^{-1}(q)]_{\mu\nu}^{\alpha\alpha} \langle j_{\nu}^{\alpha}(\vec{q})|.
\end{equation}
By time-inversion symmetry the projections on the currents are orthogonal to the densities, and the three projection operators ${\cal P}_\rho, {\cal P}_j^\parallel, {\cal P}_j^\perp$ mutually commute.
Then the Zwanzig-Mori procedure yields the second equation of motion,
\begin{align}\label{eq:eom2}
&\dot{\mathcal{K}}^{\alpha\beta}_{\mu\nu}(q,t) \nonumber \\
&+ \sum_{\kappa\lambda} \sum_{\gamma= \parallel, \perp} \int_0^t \mathfrak{M}^{\alpha\gamma}_{\mu\kappa}(q,t-t') [\bm{\mathcal{J}}^{-1}(q)]_{\kappa\lambda}^{\gamma\gamma} \mathcal{K}^{\gamma\beta}_{\lambda\nu}(q,t') \diff t' = 0,
\end{align}
where we observed again that $\bm{\mathcal{J}}(q)$ is diagonal in the channel indices. The  memory kernel $\bm{\mathfrak M}(q,t)$ of the fluctuating forces has matrix elements
\begin{equation}\label{eq:memory}
\mathfrak{M}^{\alpha\beta}_{\mu\nu}(q,t)=\frac{1}{N}\langle j_{\mu}^{\alpha}(\vec{q})|\mathcal{L}\mathcal{Q}\exp[-i\mathcal{L}_{\mathcal{Q}}t]\mathcal{Q}\mathcal{L}|j_{\nu}^{\beta}(\vec{q})\rangle,
\end{equation}
where $\mathcal{Q} = \mathcal{Q}_{j} \mathcal{Q}_{\rho} = \mathbf{1} -{\cal P}_j - {\cal P}_\rho$ projects onto the orthogonal subspace spanned by the density and the currents. The dynamics of this subspace is generated by the reduced Liouville operator $\mathcal{L}_{\mathcal{Q}}:=\mathcal{Q}\mathcal{L}\mathcal{Q}$.

The exact equations of motion assume the form of matrix-valued integro differential equations, where the memory effects emerge via the convolution integrals. We note that due to the two decay channels both integro differential equations, Eqs.~\eqref{eq:eom1} and~\eqref{eq:eom2}, with first-order derivative in time cannot be replaced by a single integro differential equation with second-order time derivative, quite in contrast to simple one and multi-component liquids.
The equations simplify in the Fourier-Laplace domain, convention
\begin{equation}
\hat{S}_{\mu\nu}(q,z)=\text{i} \int_{0}^{\infty}\diff t \, S_{\mu\nu}(q,t) \exp (\text{i} zt ), \qquad \Imag[z]>0,
\end{equation}
where $z$ constitutes a complex frequency.~\footnote{Since it is clear from the context when $z$ refers to a complex frequency or to a  distance to the wall, no confusion arises.} As usual, one infers that the Laplace transforms are analytic functions in the upper half plane $\Imag[z]>0$, and all singularities are concentrated on the complement~\cite{Goetze:Complex_Dynamics}. From the definition of the Kubo scalar product, the Laplace transforms of dynamic correlation functions are matrix elements
of the resolvent operator $({\cal L} - z)^{-1}$,  e.g.,
\begin{equation}
 \hat{S}_{\mu\nu}(q,z) = \frac{1}{N} \langle \rho_\mu(\vec{q}) |  ({\cal L}-z)^{-1} | \rho_\nu(\vec{q}) \rangle,
\end{equation}
and similarly for the other correlation functions. Transforming the first equation of motion, Eq.~\eqref{eq:eom1}, yields a matrix equation for $\hat{\bf S}(q,z)$ with formal solution,
\begin{equation}\label{eq:eom1_laplace}
 \hat{\mathbf{S}}(q,z) = -\left[ z \mathbf{S}^{-1}(q) + \mathbf{S}^{-1}(q) \hat{\mathbf{K}}(q,z)\mathbf{S}^{-1}(q) \right]^{-1}.
\end{equation}
By linearity, the decomposition of $\mathbf{K}(q,t)$ into the different relaxation channels, Eq.~\eqref{eq:split}, translates directly to the Laplace domain,
\begin{equation}\label{eq:Kcontraction}
 \hat{K}_{\mu\nu}(q,z) = \sum_{\alpha\beta=\parallel,\perp}b^{\alpha}(q,Q_{\mu}) \hat{\mathcal{K}}^{\alpha\beta}_{\mu\nu}(q,z)
 b^{\beta}(q,Q_{\nu}).
\end{equation}
Last, the second equation of motion, Eq.~\eqref{eq:eom2},  allows us to calculate the current kernel by matrix inversion
\begin{equation}\label{eq:eom2_laplace}
 \hat{\bm{\mathcal K}}(q,z) = -\left[ z \bm{\mathcal J}^{-1}(q) + \bm{\mathcal J}^{-1}(q) \hat{\bm{\mathfrak M}}(q,z)\bm{\mathcal J}^{-1}(q) \right]^{-1}.
\end{equation}
Up to this point, all equations are exact and  all  features specific to the interactions within the liquid and the wall are  encoded in the force kernel ${\bm{\mathfrak{M}}}(q,t)$.
Close to the glass transition, we anticipate that forces due to interactions persist for long times, implying that the Laplace transform  $\hat{\bm{\mathfrak{M}}}(q,z)$ becomes large for small frequencies. By Eq.~\eqref{eq:eom2_laplace} the current correlator $\hat{\bm{\mathcal{K}}}(q,z)$ becomes small in this case reflecting that transport is drastically suppressed. The first equation of motion in the Laplace domain, Eq.~\eqref{eq:eom1_laplace}, implies that the density correlation function $\hat{\mathbf{{S}}}(q,z)$ diverges for $z\to 0$ at the glass transition due to
the slowing down of the  structural relaxation.

\section{MODE-COUPLING THEORY}

The Zwanzig-Mori formalism expresses the density dynamics in terms of the force kernel ${\bm{\mathfrak{M}}}(q,t)$. To close the set of dynamic equations we need to specify the force kernel by a suitable approximation. The basic insight is that caging by neighboring particles  is the driving mechanism for the slowing down of the dynamics. Yet, the caging forces entering the force kernel are generated by the interactions with the particles, i.e., by products of density modes.  Here, we rely on the mode-coupling idea for supercooled liquids~\cite{Goetze:Complex_Dynamics} to establish a connection in the temporal domain between the force kernel as a functional of the density correlation functions. The goal is thus to derive  a microscopic theory without free parameters allowing us to evaluate the complete dynamics, including the  long-time structural relaxation, from a set of  self-consistent equations.

We implement the mode-coupling idea following the strategy  of simple bulk liquids: The forces are projected onto a set of fluctuating density-pair modes and the resulting four-point correlation function with reduced dynamics is factorized
into a product of density correlation  functions with the original dynamics. The technical procedure is to identify first an orthogonal  projection operator onto the pair fluctuating modes:
\begin{equation}
 \mathcal{P}_{\rho\rho}=\sum_{11'22'} |\delta \rho(1)\delta \rho(2)\rangle g(12;1'2') \langle \delta \rho(1') \delta\rho(2') | .
\end{equation}
Here we followed Ref.~\cite{Scheidsteger:1997} to simplify the notation by
 combining the wave vectors and mode indices into  super indices $i=(\vec{q}_{i},\mu_{i})$ and $i^{\prime}=(\vec{q}_{i}^{\prime},\mu_{i}^{\prime})$.
The matrix  $g(12;1'2')$  ensures  idempotency, $\mathcal{P}_{\rho\rho}^2 = \mathcal{P}_{\rho\rho}$, by   the  normalization condition
\begin{align}\label{eq:normalization}
 &\sum_{1^{\prime}2^{\prime}}g(1 2;1^{\prime} 2^{\prime})\langle \delta \rho(1^{\prime})^*\delta \rho(2^{\prime})^*
 \delta \rho(1^{\prime\prime})\delta\rho(2^{\prime\prime})\rangle\nonumber\\
&=\frac{1}{2}[\delta(1,1^{\prime\prime})\delta(2,2^{\prime\prime})+\delta(1,2^{\prime\prime})\delta(2,1^{\prime\prime})].
\end{align}
The essential part of the mode-coupling approximation is the factorization of the dynamical four-point correlation function into dynamical two-point correlation functions,
\begin{align}
&\langle\delta\rho(1)^*\delta\rho(2)^*\exp[-\text{i}\mathcal{L}_{\mathcal{Q}}t]\delta \rho(1^{\prime})\delta\rho(2^{\prime})\rangle\nonumber\\
&\approx N^{2}[ S(1,1^{\prime},t) S(2,2^{\prime},t)+(1^{\prime}\leftrightarrow 2^{\prime})] .
\end{align}
Specializing to $t=0$ yields an approximate factorization  of the static four-point correlation function
\begin{align}
&\langle \delta\rho(1)^*\delta\rho(2)^* \delta\rho(1^{\prime})\delta\rho(2^{\prime})\rangle\nonumber\\
&\approx \langle\rho(1) | \rho(1^{\prime})\rangle \langle \rho(2) |\rho(2^{\prime})\rangle+(1^{\prime}\leftrightarrow2^{\prime}).
\end{align}
For consistency, we employ the same factorization also in the normalization condition, Eq.~\eqref{eq:normalization}, which then allows us to determine
\begin{equation}
 g(12;1^{\prime}2^{\prime}) \approx \frac{1}{4N^{2}}\left\{ [\mathbf {S}^{-1}](1,1^{\prime})[\mathbf {S}^{-1}](2,2^{\prime})+(1^{\prime}\leftrightarrow2^{\prime})\right\}.
\end{equation}
Collecting terms the mode-coupling procedure leads to an approximation for the force kernel as a bilinear functional of the generalized intermediate scattering function,
\begin{align}\label{eq:Memory}
 [\bm{\mathfrak{M}}(q,t)]^{\alpha\beta}_{\mu\nu}&\approx  \frac{1}{2N^{3}} \sum_{\vec{q}_{1},\vec{q}_{2}=\vec{q}-\vec{q}_{1}}\sum_{\substack{\mu_{1}\mu_{2}\\ \nu_{1} \nu_{2}}}\mathcal{X}^{\alpha}_{\mu,\mu_{1}\mu_{2}}(\vec{q},\vec{q}_{1}\vec{q}_{2})\nonumber\\
&\times S_{\mu_{1}\nu_{1}}(q_{1},t) S_{\mu_{2}\nu_{2}}(q_{2},t) \mathcal{X}^{\beta}_{\nu,\nu_{1}\nu_{2}}(\vec{q},\vec{q}_{1}\vec{q}_{2})^*
\end{align}
Note that, due to translational invariance in lateral direction to the walls, only wave vectors $\vec{q}_{1}$ and $\vec{q}_{2}$ contribute which fulfill the selection rule $\vec{q}=\vec{q}_{1}+\vec{q}_{2}$.  Here the complex-valued vertices $\mathcal{X}^{\alpha}_{\mu,\mu_{1}\mu_{2}}(\vec{q},\vec{q}_{1}\vec{q}_{2})$ arise from the overlap of the fluctuating forces with the density-pair modes
\begin{align}\label{eq:vertex}
\mathcal{X}^{\alpha}_{\mu,\mu_{1}\mu_{2}}(\vec{q},\vec{q}_{1}\vec{q}_{2})=&\sum_{\mu_{1}^{\prime}\mu_{2}^{\prime}} \langle \mathcal{Q} \mathcal{L}j_{\mu}^{\alpha}(\vec{q})^*\delta \rho_{\mu_{1}^{\prime}}(\vec{q}_{1})\delta \rho_{\mu_{2}^{\prime}}(\vec{q}_{2})\rangle \nonumber\\
&\times [\mathbf{S}^{-1}(q_{1})]_{\mu_{1}'\mu_{1}} [\mathbf{S}^{-1}(q_{2})]_{\mu_{2}'\mu_{2}}.
\end{align}

The overlaps  can be evaluated explicitly in terms of structural quantities
\begin{align}\label{eq:overlap}
&\langle \mathcal{Q}\mathcal{L}j_{\mu}^{\alpha}(\vec{q})^*\delta \rho_{\mu_{1}}(\vec{q}_{1})\delta \rho_{\mu_{2}}(\vec{q}_{2})\rangle
=N\frac{k_{B}T}{m}\delta_{\vec{q},\vec{q}_{1}+\vec{q}_{2}} \nonumber \\
&\times
\Big[ b^{\alpha}(\hat{\vec{q}}\cdot\vec{q}_{1},Q_{\mu_{1}})S_{\mu-\mu_{1},\mu_2}(q_{2})+(1\leftrightarrow 2)\nonumber\\
&-\sum_{\kappa\sigma}\frac{n^*_{\mu-\kappa}}{n_{0}}b^{\alpha}(q,Q_{\kappa})[\mathbf{S}^{-1}(q)]_{\kappa\sigma}S_{\sigma,\mu_{1}\mu_{2}}(\vec{q},
\vec{q}_{1}\vec{q}_{2}) \Big] ;
\end{align}
see Appendix ~\ref{sec:scalarproduct_a}.
Here, static correlations of the density with pair modes occur, which introduces the triple correlation function
\begin{equation}\label{eq:triple}
S_{\sigma,\mu_{1}\mu_{2}}(\vec{q},\vec{q}_{1}\vec{q}_{2})=\frac{1}{N}\langle\delta\rho_{\sigma}(\vec{q})^*\delta\rho_{\mu_{1}}
(\vec{q}_{1})\delta\rho_{\mu_{2}}(\vec{q}_{2})\rangle.
\end{equation}

In practice, the static triple correlations are difficult to determine and therefore further approximations are introduced. In simple and molecular bulk liquids the convolution approximation~\cite{Goetze:Complex_Dynamics}
has proven successful to describe the glassy behavior. In Appendix ~\ref{sec:convolution} we prove that the convolution approximation applied to a liquid confined in a slit (see Appendix ~\ref{sec:convolution2} for details) leads to a similar vertex structure as found for simple and molecular liquids.
As a result the vertices assume the compact form
\begin{align}\label{eq:VertexX}
&\mathcal{X}^{\alpha}_{\mu,\mu_{1}\mu_{2}}(\vec{q},\vec{q}_{1},\vec{q}_{2})\approx -N\frac{k_{B}T}{m} \frac{n_{0}}{L^2} \delta_{\vec{q},\vec{q}_{1}+\vec{q}_{2}} \nonumber \\
&\times[b^{\alpha}(\hat{\vec{q}}\cdot\vec{q}_{1},Q_{\mu-\mu_{2}})c_{\mu-\mu_{2},\mu_{1}}(q_{1})+(1\leftrightarrow2)].
\end{align}
Here $c_{\mu\nu}(q)$ are the matrix elements of the direct correlation function implicitly defined by the  proper generalization of the Ornstein-Zernike equation,
\begin{equation}\label{eq:OZ}
 \mathbf{S}^{-1}(q)=\frac{n_{0}}{L^2}[\mathbf{v}-\mathbf{c}(q)],
\end{equation}
with $[\mathbf{v}]_{\mu\nu} = v_{\nu-\mu}$.~\footnote{In our previous work, Ref.~\cite{Lang:2010}, we considered only symmetric walls, where $[\mathbf{v}]_{\mu\nu} = v_{\nu-\mu}= v_{\mu-\nu}$.} Direct inspection shows that $c_{\mu\nu}(q)$ has the same symmetry properties as $S_{\mu\nu}(q)$.

The fluctuating force kernel enters the Zwanzig-Mori equation, Eq.~\eqref{eq:eom2_laplace}, only in terms of the combination  $\bm{\mathcal {J}}^{-1}(q)\bm{\hat{\mathfrak{M}}}(q,z)\bm{\mathcal {J}}^{-1}(q)$. This suggests to define an effective force kernel $\bm{\mathcal{M}}(q,t)$, which is then given by
\begin{align}\label{eq:effective}
&\mathcal{M}^{\alpha\beta}_{\mu\nu}(q,t)=[\bm{\mathcal {J}}^{-1}(q)\bm{\mathfrak{M}}(q,t)\bm{\mathcal {J}}^{-1}(q)]^{\alpha\beta}_{\mu\nu}\approx\mathcal{F}_{\mu\nu}^{\alpha\beta}[\mathbf{S}(t),\mathbf{S}(t);q] \nonumber\\
&=\frac{1}{2N} \sum_{\vec{q}_{1},\vec{q}_{2}=\vec{q}-\vec{q}_{1}}\sum_{\substack{\mu_{1}\mu_{2}\\ \nu_{1} \nu_{2}}}\mathcal{Y}^{\alpha}_{\mu,\mu_{1}\mu_{2}}(\vec{q},\vec{q}_{1}\vec{q}_{2})\nonumber\\
&\times S_{\mu_{1}\nu_{1}}(q_{1},t) S_{\mu_{2}\nu_{2}}(q_{2},t) \mathcal{Y}^{\beta}_{\nu,\nu_{1}\nu_{2}}(\vec{q},\vec{q}_{1}\vec{q}_{2})^*,
\end{align}
with new vertices
\begin{align}\label{eq:Ypsilon}
&\mathcal{Y}^{\alpha}_{\mu,\mu_{1}\mu_{2}}(\vec{q},\vec{q}_{1}\vec{q}_{2})\nonumber \\
&=\frac{n_0^2}{L^4}\sum\limits_{\kappa} v^*_{\mu-\kappa}[b^\alpha(\hat{\vec{q}}\cdot
\vec{q}_{1},Q_{\kappa-\mu_2}) c_{\kappa-\mu_2,\mu_1}(q_1)+(1\leftrightarrow2)].
\end{align}
Note that the static inverse current correlator, Eq.~\eqref{eq:currents_static_inverse}, is diagonal with respect to the channel index $\alpha$ but not with respect to the mode indices $\mu,\nu$.
The notation for the MCT functional $\bm{\mathcal{F}}[\mathbf{S}(t),\mathbf{S}(t);q]$ emphasizes the bilinearity
 with respect to the generalized intermediate scattering functions, which is a direct implication of the mode-coupling approximation.

The MCT equations for confined fluids closely resemble the ones for molecular liquids. This motivates us to define a class of mode-coupling theories which is distinguished  by multiple relaxation channels for current kernels.  The mathematical properties proven in the next section therefore hold not only for MCT of confined liquids but to all MCT theories belonging to this class.

\section{NONERGODICITY PARAMETER}
In this section we introduce the nonergodicity parameter, which plays a key role in glass physics for representing the spontaneous arrest of density fluctuations. It is also known as the glass form factor and  allows us to discriminate between an ergodic "liquid phase" and a non-ergodic "glass phase". The first subsection of this part provides general information about  the long-time behavior of the self-consistent set of equations for confined liquids. In the remaining  Subsections, we  prove certain mathematical aspects of the mode-coupling equations. In particular, we show that an iteration scheme can be defined where convergence to a solution for the nonergodicity parameter is ensured. Furthermore this solution is distinguished by the property that it fulfills a certain maximum principle. These last subsections  may be skipped upon the first reading of the paper.

\subsection{GENERAL DEFINITIONS}
In this section we show that the set of self-consistent equations for the intermediate scattering function can be solved for  their respective long-time limits  without solving explicitly for the dynamics for all times. Here, we adopt the same approach as for bulk systems and employ  a nonvanishing  long-time limit of the generalized intermediate scattering function
\begin{equation}
F_{\mu\nu}(q):=\lim_{t\rightarrow\infty}S_{\mu\nu}(q,t) \neq 0,
\end{equation}
as the definition for a glassy state. In the current case they constitute an infinite matrix that inherits the hermitian structure of the scattering function,
\begin{equation}
 F_{\mu\nu}(q)= F_{\nu\mu}(q)^*.
\end{equation}
For symmetric walls, the  inversion symmetry $\mathcal{I}$ yields additionally
\begin{equation}
F_{\mu\nu}(q) =F_{\mu\nu}(q)^* = F_{-\mu-\nu}(q),
\end{equation}
so the matrices are real symmetric and invariant under simultaneous change of sign of the mode indices.

We argue that on general grounds the nonergodicity parameter is a nonnegative matrix $\mathbf{F}(q) \succeq 0$. Indeed, for any set of complex numbers $y_\nu$, $\sum_{\mu \nu } y_\mu^* S_{\mu\nu}(q,t) y_\nu$ constitutes the autocorrelation function of the variables $\sum_\mu y_\mu \delta\rho_\mu(\vec{q},t) $ and as such its long-time limit is non-negative~\cite{Goetze:Complex_Dynamics}.

By the Laplace transform a nonergodic contribution results in a zero-frequency pole, $S_{\mu\nu}(q,z) = - F_{\mu\nu}(q)/z + (\text{smooth})$, for small complex frequencies $z$. Reversely, the nonergodicity parameter can be obtained from the  limit
\begin{equation}
 F_{\mu\nu}(q):=-\lim_{z\rightarrow0}z\hat{S}_{\mu\nu}(q,z).
\end{equation}
By the mode-coupling approximation, an arrest of the density modes is accompanied by a freezing of the force kernel,
\begin{equation}\label{eq:N}
 \bm{\mathcal{N}}(q) := \bm{\mathcal{M}}(q,t\to \infty ) = \bm{\mathcal{F}}[\mathbf{F},\mathbf{F};q],
\end{equation}
which is again a non-negative matrix with respect to the double index $\gamma:=(\alpha,\mu),\delta:=(\beta,\nu)$ since it is the long-time limit of an autocorrelation function. We  demonstrate in the next subsection that the MCT approximation preserves this property.
The simultaneous freezing of the forces and the densities implies again a zero-frequency pole for $\hat{M}_{\mu\nu}^{\alpha\beta}(q,z)$ and by Eqs.~\eqref{eq:eom2_laplace} and~\eqref{eq:Kcontraction}, the current correlator vanishes for small complex  frequencies $z\to 0$ as $\hat{K}_{\mu\nu}(q,z) = z G_{\mu\nu}(q)+ o(z)$ where
\begin{align}\label{eq:G}
G_{\mu\nu}(q)=& \sum_{\alpha\beta =\parallel,\perp} b^\alpha(q,Q_\mu)
[\bm{\mathcal{N}}^{-1}(q)]_{\mu\nu}^{\alpha\beta}
b^\beta(q,Q_\nu).
\end{align}
In particular, one infers $\mathbf{G}(q) \succeq 0$, since the inverse of a non-negative matrix inherits the same property as well as its contraction with respect to the channel indices. From the first equation of motion, Eq. (\ref{eq:eom1_laplace}), the nonergodicity parameter can be evaluated as
\begin{align}\label{eq:F}
 \mathbf{F}(q) =& \left[ \mathbf{S}^{-1}(q) + \mathbf{S}^{-1}(q) \mathbf{G}(q) \mathbf{S}^{-1}(q) \right]^{-1} \nonumber\\
=& \mathbf{S}(q) - \left[ \mathbf{S}^{-1}(q) + \mathbf{G}^{-1}(q)\right]^{-1}.
\end{align}
The long-time limit of the mode-coupling equations is a solution of the set of Eqs.~\eqref{eq:N}-\eqref{eq:F}. To avoid cumbersome notation
we allow $\mathbf{G}(q)$ to become formally infinite; in that case we put $\mathbf{G}^{-1}(q) =0$.
In general, these equations possess many solutions, in particular $\mathbf{F}(q) \equiv 0$ represents the  trivial solution, which corresponds to an ergodic liquid.
%Here we show that iterating the equations starting with $\mathbf{F}^{(0)}(q) = \mathbf{S}(q) \succ 0$ yields a monotone sequence $\mathbf{S}(q) \succ\mathbf{F}^{(n) }(q) \succeq \mathbf{F}^{(n+1)}(q) \succeq 0, n=1,2 \ldots$. Thus the limit $\bar{\mathbf{F}}(q) = \lim_{n\to \infty} \mathbf{F}^{(n)}(q) \succeq 0$ within the class of non-negative matrices is guaranteed.

\subsection{POSITIVITY OF THE MODE-COUPLING FUNCTIONAL}
To demonstrate the positivity property, it is convenient to introduce the pair-mode indices $a:=(\mu_{1},\mu_{2})$ and $b:=(\nu_{1},\nu_{2})$. Then Eq.~\eqref{eq:effective} allows for the compact expression
\begin{align}\label{eq:superhyper}
&\mathcal{F}[\mathbf{F},\mathbf{F};q]^{\gamma \delta}\nonumber\\
&=\frac{1}{2N}\sum_{ab}
\sum_{\vec{q}_{1},\vec{q}_{2}=\vec{q}-\vec{q}_{1}}
\mathcal{Y}^{\gamma}_{a}(\vec{q},\vec{q}_{1}\vec{q}_{2}) [\mathbf{F}(q_{1})\otimes\mathbf{F}(q_{2}) ]_{ab}\mathcal{Y}^{\delta}_{b}(\vec{q},\vec{q}_{1}\vec{q}_{2})^*,
\end{align}
where $\otimes$ denotes the Kronecker product  in the space of  mode indices, $E_{\mu_{1}\nu_{1}}(q_{1})F_{\mu_{2}\nu_{2}}(q_{2})=[\mathbf{E}(q_{1})\otimes \mathbf{F}(q_{2})]_{a=(\mu_{1},\mu_{2}),b=(\nu_{1},\nu_{2})}$.

%\begin{align}
%&\mathbf{E}(q_{1})\otimes \mathbf{F}(q_{2}):=[E_{\mu_{1}\nu_{1}}(q_{1})\cdot \mathbf{F}(q_{2})]=\nonumber\\
%&\begin{bmatrix}
%\ddots	& \dots	                               & \vdots                                     & \dots                              & \adots    \\
%        &  \ddots                              & E_{-10}(q_{1})\boldsymbol{F}(q_{2})       & \adots                             &           \\[6pt]
%\dots	& E_{0-1}(q_{1})\boldsymbol{F}(q_{2})  & E_{00}(q_{1})\boldsymbol{F}(q_{2}) 	    & E_{01}(q_{1})\boldsymbol{F}(q_{2}) & \dots     \\
%        &  \adots                              & E_{10}(q_{1})\boldsymbol{F}(q_{2})         &  \ddots                            &           \\
%\adots	& \dots	                               & \vdots                                     & \dots                              & \ddots
%\end{bmatrix}
%\end{align}

Sandwiching Eq.~\eqref{eq:superhyper} between complex-valued tuples $s^{\gamma}$ and summing over $\gamma$ yields
\begin{align}
&\sum_{\gamma \delta}s^{\gamma *} \mathcal{F}[\mathbf{F},\mathbf{F};q]^{\gamma \delta}s^{\delta}\nonumber\\
&=\frac{1}{2N}\sum_{\gamma\delta}\sum_{ab} \sum_{\vec{q}_{1},\vec{q}_{2}=\vec{q}-\vec{q}_{1}}
s^{\gamma *} \mathcal{Y}^{\gamma}_{a}(\vec{q},\vec{q}_{1}\vec{q}_{2}) [\mathbf{F}(q_{1})\otimes\mathbf{F}(q_{2}) ]_{ab}\nonumber\\
&\times\mathcal{Y}^{\delta}_{b}(\vec{q},\vec{q}_{1}\vec{q}_{2})^*s^{\delta}\nonumber\\
&=\frac{1}{2N}\sum_{ab} \sum_{\vec{q}_{1},\vec{q}_{2}=\vec{q}-\vec{q}_{1}}{\cal Z}_{a}(\vec{q},\vec{q}_{1}\vec{q}_{2})^* [\mathbf{F}(q_{1})\otimes\mathbf{F}(q_{2}) ]_{ab}\nonumber\\
&\times {\cal Z}_{b}(\vec{q},\vec{q}_{1}\vec{q}_{2}).
\end{align}
The first ingredient is the Kronecker product  of non-negative matrices which is non-negative again. Second, a contraction with complex-valued tuples ${\cal Z}_{a}(\vec{q},\vec{q}_{1}\vec{q}_{2})^*:= \sum_{\gamma}s^{\gamma *} \mathcal{Y}^{\gamma}_{a}(\vec{q},\vec{q}_{1}\vec{q}_{2})$ is performed yielding a non-negative number. Thus, the mode-coupling functional maps non-negative matrices in the mode indices $\mu,\nu$ to non-negative matrices with respect to the double indices $\gamma, \delta$ for each wave vector $q$. Generically, all vertices are nonvanishing and all components of the functional are positive matrices $\bm{\mathcal{F}}[\mathbf{F} ,\mathbf{F};q ] \succ 0$ provided the arguments are positive, $\mathbf{F}(q) \succ 0$.

\subsection{A CONVERGENT  ITERATION SCHEME}
First, we show that the mode-coupling functional  $\bm{\mathcal{N}}[\mathbf{F};q] := \bm{\mathcal{F}}[\mathbf{F},\mathbf{F};q]$ preserves the following
partial ordering: $\mathbf{F} \succeq \mathbf{E}$ if $\mathbf{F}(q)-\mathbf{E}(q) \succeq 0$ for all $q$. It is convenient to use a representation of Eq.~\eqref{eq:superhyper} that makes the symmetry upon exchanging the slots manifest,
\begin{align}
 &\mathcal{F}[\mathbf{F},\mathbf{E};q]^{\gamma \delta}=\frac{1}{4N} \sum_{ab} \sum_{\vec{q}_{1},\vec{q}_{2}=\vec{q}-\vec{q}_{1}}\mathcal{Y}^{\gamma}_{a}(\vec{q},\vec{q}_{1}\vec{q}_{2})\nonumber\\
& \times [\mathbf{F}(q_{1})\otimes\mathbf{E}(q_{2})+\mathbf{E}(q_{1})\otimes\mathbf{F}(q_{2}) ]_{ab}\mathcal{Y}^{\delta}_{b}(\vec{q},\vec{q}_{1}\vec{q}_{2})^*.
\end{align}
In the following we suppress the dependence on $q$ and all operations are to be understood componentwise for each $q$.
For given $\mathbf{F}\succeq 0$ and $\mathbf{E}\succeq 0$ the arguments for showing the positivity of the functional are easily adapted to show  $\bm{\mathcal{F}}[\mathbf{F},\mathbf{E}]\succeq 0$.
 Assuming $\mathbf{F}\succeq\mathbf{E}$ one derives  $\bm{\mathcal{N}}[\mathbf{F}]-\bm{\mathcal{N}}[\mathbf{E}]=\bm{\mathcal{F}}[\mathbf{F}+\mathbf{E},\mathbf{F}-\mathbf{E}]\succeq 0$.
Thus the mode-coupling functional preserves ordering $\bm{\mathcal{N}}[\mathbf{F}] \succeq\bm{\mathcal{N}}[\mathbf{E}]$.

Since inversion reverses ordering it follows that $\bm{\mathcal{N}}^{-1}[\mathbf{E}]-\bm{\mathcal{N}}^{-1}[\mathbf{F}]\succeq 0$ and therefore also the contractions, Eq.~\eqref{eq:G},
 fulfill $\mathbf{G}[\mathbf{E}]-\mathbf{G}[\mathbf{F}]\succeq 0$. Eventually, the mapping
\begin{equation}\label{eq:mapping}
\bm{\mathcal{I}}[\mathbf{F}] := \mathbf{S} - \left[ \mathbf{S}^{-1} + \mathbf{G}^{-1}[\mathbf{F}]\right]^{-1},
%\left[ \mathbf{S}^{-1}(q) + \mathbf{S}^{-1}(q) \mathbf{G}[\mathbf{F};q] \mathbf{S}^{-1}(q) \right]^{-1}
\end{equation}
is continuous and
also preserves the ordering
\begin{equation}
\bm{\mathcal{I}}[\mathbf{F}]-\bm{\mathcal{I}}[\mathbf{E}]\succeq 0.
\end{equation}
Since $\mathbf{G}[\mathbf{F}]\succeq 0$, positivity is inherited for the images of the mapping $\bm{\mathcal{I}}[\mathbf{F}]\succeq 0$. Furthermore $\mathbf{S} \succ \bm{\mathcal{I}}[\mathbf{F}]$
for $\mathbf{F} \succ 0$, and all fixed points $\bar{\bar{\mathbf{F}}}\succeq 0$ fulfill $\mathbf{S} \succ \bar{\bar{\mathbf{F}}}$.

We define a sequence $\mathbf{F}^{(n+1)}=\bm{\mathcal{I}}[\mathbf{F}^{(n)}]$ with initial value $\mathbf{F}^{(0)}=\mathbf{S}\succ 0$.  Since $\mathbf{G}[\mathbf{S}] \succ 0$ the first iteration leads to a matrix that is strictly smaller
 $\mathbf{S}\succ \mathbf{F}^{(1)}$. By induction one infers that the sequence is monotone and bounded
$\mathbf{S} \succ \mathbf{F}^{(n) } \succeq \mathbf{F}^{(n+1)} \succeq 0, n=1,2 \ldots$ and thus converges to some  non-negative fixed point $\bar{\mathbf{F}}\succeq 0$.

\subsection{COVARIANCE AND MAXIMUM PRINCIPLE}\label{sec:transform}
Here we show that the limit $\bar{\mathbf{F}}$ obtained by iteration with initial condition $\mathbf{F}^{(0)}=\mathbf{S}$ represents a maximal solution in the sense that all other non-negative solutions
$\bar{\bar{\mathbf{F}}} \succeq 0$ of the equation
\begin{equation}\label{eq:perl}
\mathbf{F}=\bm{\mathcal{I}}[\mathbf{F}]
\end{equation}
are smaller or equal than  $\bar{\mathbf{F}}$, i.e., $\bar{\mathbf{F}} \succeq\bar{\bar{\mathbf{F}}}$. $\bar{\mathbf{F}}$ is uniquely determined by this maximum property.
The corresponding proof is based on the covariance of Eq.~\eqref{eq:perl} under the linear transformation
\begin{equation}\label{eq:ftrans}
\bm{\mathcal{T}} : \mathbf{F}\mapsto \bm{\mathcal{T}}[\mathbf{F}]= \mathbf{F}-\bar{\bar{\mathbf{F}}}=:\tilde{\mathbf{F}},
\end{equation}
which  maps $\bar{\bar{\mathbf{F}}}$ to $0$ and $\mathbf{S}\succ 0$ to $\tilde {\mathbf{S}}=\mathbf{S}-\bar{\bar{\mathbf{F}}}\succ 0$. The latter relation follows since Eq.\eqref{eq:mapping} requires $\mathbf{S}\succ\bm{\mathcal{I}}[\mathbf{F}]$ for all $\mathbf{F} \succ 0$.  The requirement of  covariance implies that there is a transformed map $\tilde{\bm{\mathcal{I}}}[\tilde{\mathbf{F}}]$ such that
\begin{equation}\label{eq:trans}
\tilde{\mathbf{F}}=\tilde{\bm{\mathcal{I}}}[\tilde{\mathbf{F}}].
\end{equation}
The transformed functional $\tilde{\bm{\mathcal{I}}}[\tilde{\mathbf{F}}]$ is chosen such that it is linked to the original functional $\bm{\mathcal{I}}[\mathbf{F}]$ via
\begin{equation}\label{eq:def1}
\tilde{\bm{\mathcal{I}}}[\tilde{\mathbf{F}}]=\bm{\mathcal{I}}[\mathbf{F}]-\bar{\bar{\mathbf{F}}}.
\end{equation}
This requirement directly ensures that if $\mathbf{F}$ is a fixed point of $\bm{\mathcal{I}}$, then $\tilde{\mathbf{F}}$ is a fixed point of $\tilde{\bm{\mathcal{I}}}[\tilde{\mathbf{F}}]$. Thus, analogously to Eq.~\eqref{eq:mapping} we define $\tilde{\mathbf{G}}[\tilde{\mathbf{F}}]$ by

\begin{equation}\label{eq:transform}
\tilde {\bm{\mathcal{I}}}[\tilde{\mathbf{F}}] =
\tilde{\mathbf{S}} - \left[ \tilde{\mathbf{S}}^{-1} + \tilde{\mathbf{G}}^{-1}[\tilde{\mathbf{F}}] \right]^{-1}.
\end{equation}
Substituting $\bm{\mathcal{I}}[\mathbf{F}]$ from Eq.~\eqref{eq:mapping} and $\tilde {\bm{\mathcal{I}}}[\tilde{\mathbf{F}}]$ from Eq.~\eqref{eq:transform} into Eq.~\eqref{eq:def1} and taking into account that $\bm{\mathcal{I}}[\bar{\bar{\mathbf{F}}}]=\bar{\bar{\mathbf{F}}}$ we find the renormalized functional
\begin{equation}\label{eq:cofunctional}
 \tilde{\mathbf{G}}[\tilde{\mathbf{F}}]:=\left[\mathbf{G}^{-1}[\mathbf{F}]-\mathbf{G}^{-1}[\bar{\bar{\mathbf{F}}}] \right]^{-1}.
\end{equation}
For non-negative transformed functions $\tilde{\mathbf{F}}$, i.e., for $\mathbf{F}\succeq\bar{\bar{\mathbf{F}}}$, it follows from the previous subsection that $(\mathbf{G}^{-1}[\mathbf{F}]-\mathbf{G}^{-1}[\bar{\bar{\mathbf{F}}}])\succeq 0$. Consequently, the renormalized functional is again positive: $\tilde{\mathbf{G}}[\tilde{\mathbf{F}}]\succeq 0$ for $\tilde{\mathbf{F}}\succeq 0$. This in turn implies that all properties discussed in the previous subsection remain true for $\tilde {\bm{\mathcal{I}}}[\tilde{\mathbf{F}}]$ on the subspace of nonnegative $\tilde{\mathbf{F}}$, as well. Hence, the iteration of  Eq.~\eqref{eq:trans} with initial value $\tilde{\mathbf{F}}^{(0)}=\tilde {\mathbf{S}}=\mathbf{S}-\bar{\bar{\mathbf{F}}}\succ 0$ yields a fixed point  $\bar{\tilde{\mathbf{F}}}=\lim_{n\to \infty}\tilde{\mathbf{F}}^{(n)}$, which is non-negative $\bar{\tilde{\mathbf{F}}}\succeq 0$.
By construction of the transformed functional $\tilde {\bm{\mathcal{I}}}[\tilde{\mathbf{F}}]$, Eq.~\eqref{eq:def1}, the diagram
\begin{align}\label{eq:Diagram}
\begin{xy}
  \xymatrix{
      \mathbf{F} \ar[r]^{\bm{\mathcal{I}}} \ar[d]_{\bm{\mathcal{T}}}    &  \bm{\mathcal{I}}[\mathbf{F}] \ar[d]^{\bm{\mathcal{T}}}  \\
      \tilde{\mathbf{F}} \ar[r]^{\tilde{\bm{\mathcal{I}}}}             &    \tilde{\bm{\mathcal{I}}}[\tilde{\mathbf{F}}]
  }
\end{xy}
\end{align}
commutes. This property  implies  that  the sequences generated by the maps $\bm{\mathcal{I}}[\mathbf{F}]$ and $\tilde{\bm{\mathcal{I}}}[\tilde{\mathbf{F}}]$
 are in a one-to-one correspondence:
$\tilde{\mathbf{F}}^{(n)} = \mathbf{F}^{(n)} - \bar{\bar{\mathbf{F}}}$.
Thus, in the limit $n \to \infty$ the fixed points of the transformed and the original equation obey $\bar{\tilde{\mathbf{F}}}=\bar{\mathbf{F}}-\bar{\bar{\mathbf{F}}} \succeq 0$. Hence, the fixed point $\bar{\mathbf{F}}$ is larger or equal to the assumed fixed point $\bar{\bar{\mathbf{F}}}$:
\begin{equation}\label{eq:max}
\bar{\mathbf{F}}\succeq \bar{\bar{\mathbf{F}}}.
\end{equation}
Since the above argument applies to any fixed point $\bar{\bar{\mathbf{F}}}\succeq 0$, the fixed point $\bar{\mathbf{F}}$ is
the largest  non-negative solution  of the self-consistent equation determining the nonergodicity parameter. This property will be referred to as maximum principle.
Suppose now $\mathbf{F}_{*}$ is a non-negative fixed point solution which fulfills the maximum condition Eq.~\eqref{eq:max}, i.e., $\mathbf{F}_{*}\succeq \bar{\bar{\mathbf{F}}}$ for all fixed points of Eq.~\eqref{eq:perl}. Since $\bar{\mathbf{F}}$ is such a fixed point it is
\begin{equation}\label{eq:star}
\mathbf{F}_{*}\succeq \bar{\mathbf{F}}.
\end{equation}
On the other hand we can also choose in Eq.~\eqref{eq:max} $\bar{\bar{\mathbf{F}}}=\mathbf{F}_{*}$ since $\bar{\bar{\mathbf{F}}}$ is any of the fixed points, i.e., we obtain
\begin{equation}
\bar{\mathbf{F}} \succeq \mathbf{F}_{*}.
\end{equation}
Together with Eq.~\eqref{eq:star} we conclude $\bar{\mathbf{F}} = \mathbf{F}_{*}$. Consequently the maximum property determines $\bar{\mathbf{F}}$ uniquely.

Let us note that the covariance principle used here is less restrictive than for simple bulk liquids. There it could be shown that the transformed mode-coupling functional  is  again of polynomial type~\cite{Goetze:1995}.

\section{SUMMARY AND CONCLUSION}
The mode-coupling theory for liquids in confinement~\cite{Lang:2010}
constitutes a microscopic theory that is based on first principles. The scope of the theory is all two-time correlation functions, which can be measured experimentally by scattering methods, such as neutron, X-ray, or light scattering~\cite{Hansen:Theory_of_Simple_Liquids,Goetze:Complex_Dynamics}.
The same correlation functions are readily obtained by tracking all particle positions as is performed in video microscopy \cite{Nugent:2007,Eral:2009,Eral:2011} or in computer simulations~\cite{Scheidler:2000,Scheidler:2000a,Scheidler:2004,Varnik:2002,Baschnagel:2005,Varnik:2000, Teboul:2002, Fehr:1995}. Apart from being a description for liquids, the theory is also designed as a theory for the glass transition, where the structural relaxation slows down by many orders of magnitude. The strategy was first to derive a set of exact equations of motion employing the Zwanzig-Mori formalism~\cite{Goetze:Complex_Dynamics}, which introduces a memory kernel that is a functional of all the  microscopic details on the interaction of the particles among themselves and with the walls. This functional is in general unknown, and the mode-coupling idea is to consider it as a functional local in time of the intermediate scattering functions.  The  coupling coefficients are then called vertices   and are determined from structural information only.

For the case of confining parallel and flat walls, the fluctuating density field is expanded in a complete set of symmetry-adapted modes, which are continuous functions of a wave vector parallel to the planes and a discrete mode index for the Fourier expansion perpendicular. The intermediate scattering function is naturally generalized to a matrix-valued quantity with symmetry properties inherited from microscopic
considerations. The breaking of translational symmetry perpendicular to the walls implies that the container can exchange momentum with a scattering probe,  which is reflected in the nondiagonal elements of the intermediate scattering function.

A peculiarity  occurs  since the currents associated with the density fluctuations naturally split into a component parallel and perpendicular to the container walls. This requires us to modify the structure of the equations of motion from a single generalized harmonic oscillator to two coupled equations of motion with retarded friction. The same mathematical structure  also occurs in the context of molecular liquids,
where the currents consist of a translational and a reorientational part~\cite{Scheidsteger:1997,Scheidsteger:1998,Franosch:1997,Fabbian:1999}. There the splitting was necessary to ensure that the structural relaxation dynamics is independent of the moment of inertia and mass of the molecule~\cite{Schilling:2002}.

Our approach can also be  employed for different types of confinement, such that the density modes can be expanded into a complete set of geometry-adapted modes. In practice this can be achieved only for systems that display a residual symmetry such as a rectangular duct, a cylindrical shell or a spherical cavity, but in principle also for arbitrary smooth wall surfaces.

The  MCT equations for simple liquids display a series of mathematical properties, which have been demonstrated rigorously~\cite{Goetze:1995,Goetze:Complex_Dynamics}. For example, the nonergodicity parameters can be obtained as the limit of a simple iteration scheme, which is guaranteed to converge to a non-negative solution, as required by general properties of autocorrelation functions. This solution is distinguished by a maximum principle, which is obtained by a covariance property of the set of MCT equations of motion~\cite{Goetze:1995,Goetze:Complex_Dynamics}. For multicomponent liquids the intermediate scattering function is generalized to a matrix-valued quantity where the matrix indices refer to the different species in the liquid. The ideas of the proofs can then be transferred~\cite{Franosch:2002} provided the notion of positivity is generalized to hermitian matrices with positive eigenvalues.
The  nonergodicity parameters of the MCT for confined liquids are solutions of matrix-valued equations with a sophisticated mathematical structure due to the splitting of the currents. Here we  have shown that the mode-coupling functional is positive in the matrix sense with respect to certain superindices. Then we suggested an iteration scheme for the nonergodicity parameters with initial value $\mathbf{F}^{(0)}=\mathbf{S}$, which is monotonic thereby ensuring convergence. The solution $\bar{\mathbf{F}}$ thus obtained is non-negative and fulfills a generalized maximum principle. The key again was to show that the structure of the equations determining the long-time limit  reflect a covariance property with respect to suitable    shifts of  the nonergodicity parameters. The proofs developed here readily transfer also to the case of molecular liquids, which display the same mathematical structure. Hence, our paper entails important conclusions also for the well-established mode-coupling approach for molecular liquids~\cite{Scheidsteger:1997,Scheidsteger:1998,Franosch:1997,Fabbian:1999,Letz:1999,Theis:1998,Letz:1999}  and it is encouraging to investigate MCT extensions for even more complex systems. The mathematical implications  of our work demonstrate  the robustness of the
Zwanzig-Mori procedure combined with the first-principles MCT approach. We are confident that also the mathematical properties of the dynamic equations of motion from the bulk MCT can be generalized to the case of confined liquids. In particular, one should prove the existence and uniqueness of the time-dependent solution and demonstrate that for overdamped motion the solutions correspond to pure relaxations described by a superposition of decaying exponentials only. Furthermore, the long-time limit of the intermediate scattering function is expected to coincide with the maximal solution obtained by our iteration scheme. Similarly,  we anticipate that all glass transition singularities in the MCT for confined systems are of the $A_\ell$ type, specified by the classification of Arnol'd~\cite{Arnold:1975}.

Recently, striking correlations between diffusivities of colloidal spheres in confinement with local packing properties have been observed and quantified in a series of empirical scaling properties~\cite{Goel:2009}. In particular, the mobility displays oscillations as a function of the wall separation, which is attributed to commensurability effects of the packing in confinement~\cite{Mittal:2008}. Since the mode-coupling theory for confined liquids incorporates packing effects in terms of generalized static structure factors, it appears promising that our theory constitutes a microscopic basis for the observed empirical correlations.

The confining walls induce strong anisotropic correlations in the liquid ~\cite{Nygard:2012} and cannot be treated by perturbation theory. In particular, the changes cannot be obtained as linear response to an external potential as has been investigated in Refs. ~\cite{Biroli:2006,Szamel:2010}.
Our setup requires us to consider symmetry-adapted modes from the very beginning such that the layering and local packing is incorporated in suitable static quantities.

Confinement of a liquid can also be achieved inside of a porous matrix where a glass transition can occur within the frozen structure~\cite{Krakoviack:2005,Krakoviack:2007,Krakoviack:2009,Krakoviack:2011}. In addition to the slowing down due to caging the interaction with the disordered environment can lead to a localization phenomenon. In contrast to flat parallel walls, the disordered obstacles imply an additional relaxation channel in the memory kernel, which in mode-coupling approximation results in a linear coupling to the intermediate scattering function. Such a linear coupling is expected also for the case of rough walls, where corrugations open the possibility to exchange momentum also in the parallel direction of the walls.

The mode-coupling theory for confined liquids is a microscopic theory that does not require parameter adjustments. Hence the theory can be tested by computer simulations and experiments. The required input is the static structure which is assumed to be known. The MCT equations involve the three-point static correlation function, which is typically difficult to determine.
Applying a  static convolution approximation for inhomogeneous liquids~\cite{Rajan:1978} to  slit geometry the vertices assume the same compact form as found for simple and molecular liquids.
In particular, this approach  reduces to the standard convolution approximation in the limit of bulk and two-dimensional liquids,
%i.e. in the limit $L\to\infty$ and $L\to 0$,
respectively.
While for three-dimensional homogeneous systems the convolution approximation  has been proven sufficient to capture the key features of supercooled simple liquids, see  Ref.~\cite{Sciortino:2001} for an exception, it remains a challenge for the future to clarify the quality of these different approaches.

The most promising route for experimental tests are dense colloidal suspensions confined by glass plates~\cite{Nugent:2007} where the effects of commensurability can  conveniently be studied. Our theory is applicable also for these overdamped systems provided the equations of motion are supplemented by friction terms accounting for the interaction with the solvent and dropping the inertial terms. A more rigorous approach would rely on the Smoluchowski operator in the first place  and introduce suitable one-particle irreducible memory kernels as has been done for bulk liquids~\cite{Szamel:1991}.
These modifications affect only the short-time behavior; the structural relaxation encoded in the MCT memory kernels remains unchanged~\cite{Franosch:1998a}. In particular, the phase diagram and the characteristic nonergodicity parameters are identical for atomic liquids and colloidal suspensions.

The MCT approach for the collective dynamics of confined liquids can be adapted to the case of tagged-particle motion, which is of particular interest since the self-dynamics is readily accessible in computer simulation and single-particle tracking methods on experimental samples. In particular, the incoherent  nonergodicity parameters are obtainable from a similar set of self-consistent matrix equations as for the collective ones, where the mode-coupling functional now couples to both the coherent and incoherent motion~\cite{Lang:2012}. Similarly, it would be interesting to study also the motion of the transverse currents and discuss the emergence of more than one viscosity due to the breaking of translational symmetry.

Our equations allow for a direct generalization to the case of multicomponent mixtures, which is of particular interest, since they can be easily driven to a glassy state. For confinement it is even more important to suppress the nucleation of crystals since flat walls tend to facilitate the formation of ordered structures. Mixing effects~\cite{Hajnal:2009,Voigtmann:2003,Voigtmann:2011,Weysser:2011} arise due to the presence of a new length scale characterizing the near order. In confinement this local packing  competes with the layering induced by the walls and an even richer phenomenology is expected.

\begin{acknowledgments}
It is a pleasure to thank W. G\"otze for insightful discussions on the mathematical properties of mode-coupling equations and the glass transition singularity as well as critical comments on the manuscript.
This work has been supported by the
Deutsche Forschungsgemeinschaft DFG via the  Research Unit FOR1394 ``Nonlinear Response to
Probe Vitrification.'' S.L. gratefully acknowledges the support by the Cluster of Excellence ``Engineering of Advanced Materials''
 at the University of Erlangen-Nuremberg, which is funded by the DFG
 within the framework of its ``Excellence Initiative.''
\end{acknowledgments}

\appendix

\section{Static current density correlator}
\label{sec:Current}

The static current density correlation $ \mathcal{J}_{\mu \nu}^{\alpha \beta}(q)= N^{-1}\langle  j_{\mu}^{\alpha}(\vec{q})| j_{\nu}^{\beta}(\vec{q})\rangle$
is a diagonal matrix with respect to $\alpha$ and $\beta$ as averages over unpaired  momenta, e.g., $P_{n}^{x}P_{n}^{y}$, vanish. Inserting the current densities and pre-averaging over the momenta  one obtains
\begin{align}
\mathcal{J}_{\mu \nu}^{\alpha \beta}(q)&= \frac{1}{N m^{2}}\delta_{\alpha\beta}  \sum_{n,m=1}^{N} b^{\alpha}\left(\left\langle (\hat{\vec{q}}\cdot\vec{P}_{n})(\hat{\vec{q}}\cdot{\vec{P}_{m}})\right\rangle,\left\langle P_{n}^{z}P_{m}^{z} \right\rangle\right)\nonumber \\
&\times \langle \text{e}^{\text{i} \vec{q} \cdot (\vec{r}_{m}-\vec{r}_{n})} \exp\left(\text{i} Q_{\nu}z_{m}\right) \exp\left( -\text{i}Q_{\mu}z_{n}\right)\rangle.
\end{align}
Direct evaluation of the averages over the momenta yields
\begin{equation}\label{eq:momenta}
 \langle P_{n}^z P_{m}^z\rangle=\langle (\hat{\vec{q}}\cdot\vec{P}_{n})(\hat{\vec{q}}\cdot{\vec{P}_{m}})\rangle = \delta_{nm} \, mk_{B}T,
\end{equation}
and  with $\langle \rho_\mu(\vec{q},t) \rangle = A n_\mu \delta_{\vec{q},\vec{0}}$ one obtains  the explicit expression
\begin{equation}
 \mathcal{J}_{\mu \nu}^{\alpha \beta}(q)
%=\frac{1}{N} \frac{k_{B}T}{m}\langle \rho_{\mu-\nu}(\vec{0})^*\rangle\delta_{\alpha\beta}
= \frac{k_{B}T}{m}\frac {n_{\mu-\nu}^*}{n_{0}}\delta_{\alpha\beta}.
\end{equation}

\section{Time-evolution operator identity}
\label{sec:operator-identity}
The backwards-time evolution operator $\mathcal{R}(t)=\exp(- \text{i}{\cal L}t)$  allows for the  decomposition $\mathcal{R}(t)=\mathcal{R}_{\mathcal{P}}(t)+\mathcal{R}_{\mathcal{Q}}(t)$ with $\mathcal{R}_{\mathcal{P}}(t)=\mathcal{P}\mathcal{R}(t)$ and $\mathcal{R}_{\mathcal{Q}}(t)=\mathcal{Q}\mathcal{R}(t)$. By the equation of motion  $\partial_{t}\mathcal{R}(t)=-\text{i}\mathcal{L} \mathcal{R}(t)$, one obtains
\begin{equation}
 \partial_{t}\mathcal{R}_{\mathcal{Q}}(t)=-\text{i}\mathcal{Q}\mathcal{L}\mathcal{R}_{\mathcal{P}}(t) -\text{i}\mathcal{Q}\mathcal{L}\mathcal{R}_{\mathcal{Q}}(t),
\end{equation}
which is formally solved by
\begin{equation}
\mathcal{R}_{\mathcal{Q}}(t)= \text{e}^{- \text{i}{\cal Q}{\cal L}t }{\cal Q} -\text{i} \int_{0}^{t} \text{e}^{- \text{i}{\cal Q} {\cal L} (t-t')} {\cal Q } {\cal L} \mathcal{R_{\mathcal{P}}}(t')  \diff t' .
\end{equation}
Hence  the backwards-time evolution operator can be expressed as
\begin{align}
&\mathcal{R}(t)=\mathcal{P}\mathcal{R}(t)+ \text{e}^{- \text{i}{\cal Q}{\cal L}t}{\cal Q} %\nonumber  &\\
-\text{i} \int_{0}^{t}  \text{e}^{- \text{i}{\cal Q} {\cal L} (t-t')} {\cal Q } {\cal L} \mathcal{R_{\mathcal{P}}}(t')\diff t' .
\end{align}
The reduced backwards-time evolution operator can be cast in the explicitly symmetric form $\exp(- \text{i}{\cal Q}{\cal L}t)\mathcal{Q}=\mathcal{Q}\exp(- \text{i}{\cal Q}{\cal L}\mathcal{Q}t)\mathcal{Q}$.
Multiplying the previous equation from the right by $\mathcal{P}$ and from the left by $\mathcal{P}\mathcal{L}$ one arrives at
\begin{align}
&\mathcal{P}\mathcal{L}\mathcal{R}(t)\mathcal{P}=\mathcal{P}\mathcal{L}\mathcal{P}\mathcal{R}(t)\mathcal{P}\nonumber\\
&-\text{i}\int_{0}^{t} \diff t'\mathcal{P}\mathcal{L}\mathcal{Q}  \text{e}^{- \text{i}{\cal Q} {\cal L}\mathcal{Q} (t-t')} {\cal Q } {\cal L}\mathcal{P}\mathcal{R}(t')\mathcal{P}.
\end{align}
Last, employing the equation of motion $\partial_{t} \mathcal{R}(t)=-\text{i}\mathcal{L}\mathcal{R}(t)$, the operator  identity
\begin{align}
&\partial_{t} \mathcal{P}\mathcal{R}(t)\mathcal{P}+\text{i}\mathcal{P}\mathcal{L}\mathcal{P}\mathcal{R}(t)\mathcal{P}\nonumber \\
&+\int_{0}^{t} \diff t'\mathcal{P}\mathcal{L}\mathcal{Q}  \text{e}^{- \text{i}{\cal Q} {\cal L}\mathcal{Q} (t-t')} {\cal Q } {\cal L}\mathcal{P}\mathcal{R}(t')\mathcal{P}=0
\end{align}
follows, which is the starting point of the  Zwanzig-Mori procedure.

\section{Evaluation of the overlap matrix element}\label{sec:scalarproduct_a}
Here we calculate the scalar product
$\langle \mathcal{Q} \mathcal{L}j_{\mu}^{\alpha}(\vec{q})^*\delta\rho_{\mu_1}(\vec{q_{1}})\delta\rho_{\mu_2}(\vec{q}_2)\rangle$ required for the mode-coupling vertex in Eq.~(\ref{eq:overlap}).
With $\mathcal{Q}=1-\mathcal{P}_{j}-\mathcal{P}_{\rho}$ and $\mathcal{P}_{j}|\delta\rho_{\mu_1}(\vec{q_{1}})\delta\rho_{\mu_2}(\vec{q}_2)\rangle=0$ by time inversion symmetry, one obtains three contributions:
\begin{align}\label{eq:three}
&\langle \mathcal{Q} \mathcal{L}j_{\mu}^{\alpha}(\vec{q})^*\delta\rho_{\mu_1}(\vec{q_{1}})\delta\rho_{\mu_2}(\vec{q}_2)\rangle\nonumber \\
=&\langle j_{\mu}^{\alpha}(\vec{q})^* [\mathcal{L}\delta\rho_{\mu_1}(\vec{q_{1}})]\delta\rho_{\mu_2}(\vec{q}_2)\rangle+(1\leftrightarrow2)\nonumber \\
&-\langle\mathcal{L}j_{\mu}^{\alpha}(\vec{q})^*\mathcal{P}_{\rho} [\delta \rho_{\mu_1}(\vec{q_{1}})\delta\rho_{\mu_2}(\vec{q}_2)] \rangle.
\end{align}
For the  first term the particle conservation law  Eq.~(\ref{eq:Continuity}) implies
\begin{align}\label{eq:scalar}
& \langle j_{\mu}^{\alpha}(\vec{q})^* [\mathcal{L}\delta\rho_{\mu_1}(\vec{q_{1}})]\delta\rho_{\mu_2}(\vec{q}_2)\rangle\nonumber\\
&=\sum_{\gamma}b^{\gamma}(q_1,Q_{\mu_1})\langle j_{\mu}^{\alpha}(\vec{q})^*j_{\mu_{1}}^{\gamma}(\vec{q}_{1})\delta \rho_{\mu_{2}}(\vec{q}_{2})\rangle.
\end{align}
Again, averaging over the momenta first, and then over the positions similar to Eq.~(\ref{eq:momenta}),
one obtains
\begin{align}
&\langle j_{\mu}^{\alpha}(\vec{q})^*j_{\mu_{1}}^{\gamma}(\vec{q}_{1}) \delta\rho_{\mu_{2}}(\vec{q}_{2})\rangle\nonumber\\
&=\delta_{\alpha\gamma}\delta_{\vec{q},\vec{q}_{1}+\vec{q}_{2}} \frac{k_{B}T}{m}b^\alpha(\hat{\vec{q}}\cdot\hat{\vec{q}}_{1},1)\langle\rho_{\mu-\mu_{1}}(\vec{q}_{2})|\rho_{\mu_{2}}(\vec{q}_{2})\rangle\nonumber\\
&=\delta_{\alpha\gamma}\delta_{\vec{q},\vec{q}_{1}+\vec{q}_{2}} N\frac{k_{B}T}{m}b^\alpha(\hat{\vec{q}}\cdot\hat{\vec{q}}_{1},1)S_{\mu-\mu_{1},\mu_{2}}(q_{2}).
\end{align}
Here, translational invariance implies conservation of momentum parallel to the walls $\vec{q}=\vec{q}_{1}+\vec{q}_{2}$.

Evaluating the projection on the density modes in  the third term in Eq.~(\ref{eq:three})   leads to
\begin{align}
&\langle\mathcal{L}j_{\mu}^{\alpha}(\vec{q})^* {\cal P}_{\rho}
[ \delta\rho_{\mu_1}(\vec{q_{1}})\delta\rho_{\mu_2}(\vec{q}_2)
]
\rangle \nonumber\\
=&\frac{1}{N}\sum_{\kappa,\sigma}\langle j_{\mu}^{\alpha}(\vec{q})|\mathcal{L}\rho_{\kappa}(\vec{q})\rangle [\mathbf{S}^{-1}(q)]_{\kappa\sigma} \nonumber\\
&\times\langle\delta\rho_{\sigma}(\vec{q})^*\delta\rho_{\mu_{1}}
(\vec{q}_{1})\delta\rho_{\mu_{2}}(\vec{q}_{2})\rangle\nonumber\\
=&\frac{1}{N}\sum_{\kappa,\sigma,\beta}b^{\beta}(q,Q_{\kappa})\langle j_{\mu}^{\alpha}(\vec{q})|j_{\kappa}^{\beta}(\vec{q})\rangle [\mathbf{S}^{-1}(q)]_{\kappa\sigma} \nonumber\\
&\times\langle\delta\rho_{\sigma}(\vec{q})^*\delta\rho_{\mu_{1}}
(\vec{q}_{1})\delta\rho_{\mu_{2}}(\vec{q}_{2})\rangle,
\end{align}
where  particle conservation, Eq.~(\ref{eq:Continuity}), has been used again. Substituting Eq.~(\ref{eq:currents_static}) for the current-current static correlator
  the projected matrix element evaluates to
\begin{align}
&\langle\mathcal{L}j_{\mu}^{\alpha}(\vec{q})|\mathcal{P}_{\rho}|\delta\rho_{\mu_1}(\vec{q_{1}})\delta\rho_{\mu_2}(\vec{q}_2)\rangle\nonumber\\
=&\delta_{\vec{q},\vec{q}_{1}+\vec{q}_{2}}N\frac{k_{B}T}{m}\sum_{\kappa,\sigma}\frac{n^{*}_{\mu-\kappa}}{n_0}b^\alpha(q,Q_{\kappa})\nonumber\\
&\times[\mathbf{S}^{-1}(q)]_{\kappa\sigma}S_{\sigma,\mu_{1}\mu_{2}}(\vec{q},\vec{q}_{1}\vec{q}_{2}).
\end{align}

Here, we abbreviated
the static three-point correlation function by $S_{\sigma,\mu_{1}\mu_{2}}(\vec{q},\vec{q}_{1}\vec{q}_{2})= N^{-1} \langle\delta\rho_{\sigma}(\vec{q})^*\delta\rho_{\mu_{1}}
(\vec{q}_{1})\delta\rho_{\mu_{2}}(\vec{q}_{2})\rangle$, Eq.~(\ref{eq:triple}).
Collecting terms one finds Eq.~(\ref{eq:overlap}) of the main text:
\begin{align}
&\langle\mathcal{Q} \mathcal{L}j_{\mu}^{\alpha}(\vec{q})|\delta\rho_{\mu_1}(\vec{q}_{1})\delta\rho_{\mu_2}(\vec{q}_2)\rangle\nonumber\\
=&N\frac{k_{B}T}{m}\delta_{\vec{q},\vec{q}_{1}+\vec{q}_{2}}\big\{b^\alpha( \hat{\vec{q}}\cdot\vec{q}_{1},Q_{\mu_{1}})
S_{\mu-\mu_{1},\mu_{2}}(q_{2})+(1\leftrightarrow2)\nonumber\\
&-\frac{1}{n_{0}}\sum_{\kappa,\sigma}n^*_{\mu-\kappa}b^\alpha(q,Q_{\kappa}) [\mathbf{S}^{-1}(q)]_{\kappa\sigma} S_{\sigma,\mu_{1}\mu_{2}}(\vec{q},\vec{q}_{1}\vec{q}_{2})\big\}.
\end{align}

\section{Vertex approximation}
\label{sec:convolution}

In this appendix, we complete the calculation of the MCT vertex, using
the convolution approximation in order to express the static
three-point correlation function in terms of products of two-point
correlation functions.

The vertex after evaluating the overlap matrix elements is given by
three terms [cf. Eq.~(\ref{eq:overlap})]:
\begin{multline}\label{eq:vertexinit}
  \mathcal{X}^{\alpha}_{\mu,\mu_{1}\mu_{2}}(\vec{q},\vec{q}_{1}\vec{q}_{2})
  = N \frac{k_{B}T}{m}\delta_{\vec{q},\vec{q}_{1}+\vec{q}_{2}} \Big\{
  \sum_{\mu_{1}^{\prime}\mu_{2}^{\prime}}
  \Big[b^\alpha(\hat{\vec{q}}\cdot\vec{q}_{1},Q_{\mu_{1}^{\prime}}) \\
  \times S_{\mu-\mu_{1}^{\prime},\mu_{2}^{\prime}}(q_{2})
  [\mathbf{S}^{-1}(q_{1})]_{\mu_{1}^{\prime}\mu_{1}}
  [\mathbf{S}^{-1}(q_{2})]_{\mu_{2}^{\prime}\mu_{2}} +
  (1\leftrightarrow2)\Big] \\
  - \sum_{\kappa,\sigma} \sum_{\mu_{1}^{\prime}\mu_{2}^{\prime}}
  \frac{n^*_{\mu-\kappa}}{n_{0}} b^\alpha(q,Q_{\kappa})
  [\mathbf{S}^{-1}(q)]_{\kappa\sigma} \\
  \times
  S_{\sigma,\mu_{1}^{\prime}\mu_{2}^{\prime}}(\vec{q},\vec{q}_{1}\vec{q}_{2})
  [\mathbf{S}^{-1}(q_{1})]_{\mu_{1}^{\prime}\mu_{1}}
  [\mathbf{S}^{-1}(q_{2})]_{\mu_{2}^{\prime}\mu_{2}} \Big\}.
\end{multline}
For the first two terms in the bracket, the sums over
$(\mu_{1}^\prime,\mu_{2}^\prime)$ can be performed which leads to
\begin{multline}
  b^\alpha(\hat{\vec{q}}\cdot\vec{q}_{1},Q_{\mu-\mu_{2}})
  [\mathbf{S}^{-1}(q_{1})]_{\mu-\mu_{2},\mu_{1}} \\
  + b^\alpha(\hat{\vec{q}}\cdot\vec{q}_{2},Q_{\mu-\mu_{1}})
  [\mathbf{S}^{-1}(q_{2})]_{\mu-\mu_{1},\mu_{2}}.
\end{multline}
Inserting the Ornstein-Zernike equation, Eq.~\eqref{eq:OZ}, they can
be recast to
\begin{multline}\label{eq:firsttwoterms}
  \frac{n_{0}}{L^2} \Big[ b^\alpha(q,Q_{2\mu-\mu_{1}-\mu_{2}})
  v^*_{\mu-\mu_{1}-\mu_{2}} \\
  - b^\alpha(\hat{\vec{q}}\cdot\vec{q}_{1},Q_{\mu-\mu_{2}})
  c_{\mu-\mu_{2},\mu_1}(q_{1}) \\
  - b^\alpha(\hat{\vec{q}}\cdot\vec{q}_{2},Q_{\mu-\mu_{1}})
  c_{\mu-\mu_{1},\mu_2}(q_{2}) \Big],
\end{multline}
where  the linearity of  the selector $b^{\alpha}$ for $\alpha=\parallel$ and
the selection rule $\vec{q}=\vec{q}_{1}+\vec{q}_{2}$
has been used.
As for the third term, the convolution approximation (see Appendix
\ref{sec:convolution2}) gives for the triplet structure factor
\begin{multline}
  S_{\sigma,\mu_{1}^{\prime}\mu_{2}^{\prime}}
  ({\vec{q},\vec{q}_{1}\vec{q}_{2}}) \approx
  \frac{n_0^2}{L^6} \sum_{\substack{\beta_1,\beta_2,\beta_3 \\
      \lambda_1,\lambda_2,\lambda_3}} n_{\beta_1+\beta_2+\beta_3}
  v_{-\beta_1-\lambda_1} S_{\sigma\lambda_1}(q) \\
  \times v_{-\beta_2-\lambda_2} S_{(-\lambda_2)\mu_{1}^{\prime}}(q_1)
  v_{-\beta_3-\lambda_3} S_{(-\lambda_3)\mu_{2}^{\prime}}(q_2),
\end{multline}
where we omit a redundant $\delta_{\vec{q},\vec{q}_1+\vec{q}_2}$
prefactor. One can then successively sum out
$(\mu_{1}^\prime,\mu_{2}^\prime,\sigma)$,
$(\lambda_1,\lambda_2,\lambda_3)$, $\beta_1$, and $\kappa$, to reduce
this term to
\begin{equation}
  - \frac{n_0}{L^4} \sum_{\beta_2,\beta_3} n^*_{\mu-\beta_2-\beta_3}
  b^\alpha(q,Q_{\beta_2+\beta_3}) v_{\mu_{1}-\beta_2}
  v_{\mu_{2}-\beta_3}.
\end{equation}
Further progress is made by making explicit the action of the
selector and by using the linearity of $Q_{\beta}$ with respect to its
index in the case $\alpha=\perp$. Eventually, performing the last
summations over $\beta_2$ and $\beta_3$, the third term in
Eq.~\eqref{eq:vertexinit} reduces to
\begin{equation}
  - \frac{n_0}{L^2} b^\alpha(q,Q_{2\mu-\mu_{1}-\mu_{2}})
  v^*_{\mu-\mu_{1}-\mu_{2}}
\end{equation}
and is found to cancel the first term in Eq.~\eqref{eq:firsttwoterms}.

The vertex thus simplifies to
\begin{multline}
  \mathcal{X}^{\alpha}_{\mu,\mu_{1}\mu_{2}}(\vec{q},\vec{q}_{1}\vec{q}_{2})
  \\
  \approx -N\frac{k_{B}T}{m} \delta_{\vec{q},\vec{q}_{1}+\vec{q}_{2}}
  \frac{n_{0}}{L^2}
  b^\alpha(\hat{\vec{q}}\cdot\vec{q}_{1},Q_{\mu-\mu_{2}})
  c_{\mu-\mu_{2},\mu_{1}}(q_{1}) \\
  -N\frac{k_{B}T}{m} \delta_{\vec{q},\vec{q}_{1}+\vec{q}_{2}}
  \frac{n_{0}}{L^2}
  b^\alpha(\hat{\vec{q}}\cdot\vec{q}_{2},Q_{\mu-\mu_{1}})
  c_{\mu-\mu_{1},\mu_{2}}(q_{2}),
\end{multline}
which has the same form as for simple \cite{Goetze:Complex_Dynamics}
and molecular liquids \cite{Scheidsteger:1997,Fabbian:1999}.

\section{Convolution approximation}
\label{sec:convolution2}

In this appendix based on Ref.~\cite{Rajan:1978}, we report the
expression of the triplet structure factor of an inhomogeneous fluid
system provided by the convolution approximation. We first discuss the
\textit{general} case, then specialize the equations to the slab geometry.

Consider an inhomogeneous $N$-particle fluid system enclosed in a
rectangular box of volume $V$. Its one-body density and its total
correlation function are denoted by $n(\vec{r})$ and
$h(\vec{r}_{1},\vec{r}_{2})$, respectively, with the corresponding
Fourier transforms,
\begin{gather}
  \tilde{n}(\vec{k}) = \int n(\vec{r}) e^{i\vec{k}\cdot\vec{r}}
  \diff\vec{r}, \\
  \tilde{h}(\vec{k}_{1},\vec{k}_{2}) = \int h(\vec{r}_{1},\vec{r}_{2})
  e^{i(\vec{k}_1\cdot\vec{r}_1+\vec{k}_2\cdot\vec{r}_2)}
  \diff\vec{r}_1 \diff\vec{r}_2.
\end{gather}
The triplet structure factor is defined as
\begin{equation}
  S^{(3)}({\vec{k}_1,\vec{k}_{2},\vec{k}_{3}}) = \frac{1}{N} \langle
  \delta\rho(\vec{k}_1) \delta\rho(\vec{k}_{2})
  \delta\rho(\vec{k}_{3}) \rangle,
\end{equation}
with
\begin{equation}
  \delta\rho(\vec{k}) = \rho(\vec{k}) - \langle \rho(\vec{k})
  \rangle = \rho(\vec{k}) - \tilde{n}(\vec{k})
\end{equation}
and
\begin{equation}
  \rho(\vec{k}) = \sum_{j=1}^N e^{i\vec{k}\cdot\vec{x}_j},
\end{equation}
where $\vec{x}_j$ is the position of the $j$th particle. Note that,
at variance with the main text, the definitions of the structure
factors in this appendix do not involve any complex conjugation to
preserve the symmetry of the working equations.

Following Rajan \emph{et al.} \cite{Rajan:1978}, the convolution
approximation for $S^{(3)}({\vec{k}_1,\vec{k}_{2},\vec{k}_{3}})$ reads
\begin{multline} \label{eq:Rajan}
  S^{(3)}({\vec{k}_1,\vec{k}_{2},\vec{k}_{3}}) \approx \frac{1}{N V^6}
  \sum_{\substack{\vec{K}_1,\vec{K}_2,\vec{K}_3\\
      \vec{p}_1,\vec{p}_2,\vec{p}_3}}
  \tilde{n}(\vec{p}_1 + \vec{p}_2 + \vec{p}_3) \\
  \times \prod_{i=1}^3 \tilde{n}(\vec{k}_i+\vec{K}_i) \left[
    \tilde{h}(-\vec{p}_i, -\vec{K}_i) - \tilde{h}(\vec{k}_i-\vec{p}_i,
    - \vec{k}_i-\vec{K}_i ) \right] .
\end{multline}
This result is most conveniently reformulated in terms of the pair
structure factor
\begin{equation}
  S^{(2)}({\vec{k}_1,\vec{k}_{2}}) = \frac{1}{N} \langle
    \delta\rho(\vec{k}_1) \delta\rho(\vec{k}_{2}) \rangle,
\end{equation}
related to $\tilde{h}(\vec{k}_{1},\vec{k}_{2})$ through
\begin{multline} \label{eq:HtoS} %
  N S^{(2)}({\vec{k}_1,\vec{k}_{2}}) = \tilde{n}(\vec{k}_1+\vec{k}_{2})
   \\ +\frac{1}{V^2} \sum_{\vec{p}_1,\vec{p}_2}
  \tilde{n}(\vec{k}_1-\vec{p}_1) \tilde{n}(\vec{k}_2-\vec{p}_2)
  \tilde{h}(\vec{p}_1,\vec{p}_2).
\end{multline}
Defining the local specific volume $v(\vec{r}) = 1/n(\vec{r})$ and its
Fourier transform $\tilde{v}(\vec{k})$ such that
\begin{equation}
  \frac{1}{V} \sum_{\vec{p}} \tilde{n}(\vec{k}_1-\vec{p})
  \tilde{v}(\vec{p}-\vec{k}_2) = V \delta_{\vec{k}_1,\vec{k}_2},
\end{equation}
Eq.~\eqref{eq:HtoS} is easily inverted to yield
\begin{multline} \label{eq:StoH} %
  \tilde{h}(\vec{k}_{1},\vec{k}_{2}) = - \tilde{v}(\vec{k}_1+\vec{k}_{2})
   \\ +\frac{N}{V^2} \sum_{\vec{p}_1,\vec{p}_2}
  \tilde{v}(\vec{k}_1-\vec{p}_1) \tilde{v}(\vec{k}_2-\vec{p}_2)
  S^{(2)}(\vec{p}_1,\vec{p}_2),
\end{multline}
which can be injected into Eq.~\eqref{eq:Rajan}. The summations over $
\vec{K}_1$, $\vec{K}_2$, and $\vec{K}_3$, can then be explicitly
performed and, using the fact that for a closed system (e.g. canonical system)
$S^{(2)}(\vec{k},\vec{0}) = 0$ for any
$\vec{k}$, it follows that in the convolution approximation
\begin{multline} \label{eq:convolution} %
  S^{(3)}({\vec{k}_1,\vec{k}_{2},\vec{k}_{3}}) \approx \frac{N^2}{V^6}
  \sum_{\substack{\vec{p}_1,\vec{p}_2,\vec{p}_3 \\
      \vec{l}_1,\vec{l}_2,\vec{l}_3}}
  \tilde{n}(\vec{p}_1 + \vec{p}_2 + \vec{p}_3) \\
  \times \prod_{i=1}^3 \tilde{v}(-\vec{p}_i-\vec{l}_i)
  S^{(2)}(\vec{l}_i,\vec{k}_i) .
\end{multline}

One can readily check that this expression reproduces the standard
result for bulk systems. Indeed, one then has $\tilde{n}(\vec{k})=N
\delta_{\vec{k},\vec{0}}$, $\tilde{v}(\vec{k})= (V^2/N)
\delta_{\vec{k},\vec{0}}$, and $S^{(2)}(\vec{p},\vec{k}) = S(k)
\delta_{\vec{p}+\vec{k},\vec{0}}$, so that
\begin{equation}
  S^{(3)}({\vec{k}_1,\vec{k}_{2},\vec{k}_{3}}) \approx
  \delta_{\vec{k}_1+\vec{k}_2+\vec{k}_3,\vec{0}} S(k_1) S(k_2) S(k_3).
\end{equation}

Application to the slab geometry is just as straightforward. One
simply has to set $V=L A$, split each sum over a wave vector $\vec{k}=(\vec{q},Q_{\mu})$ into one
over a transverse index $\mu$  and one over an in-plane wave vector $\vec{q}$, and
replace $\tilde{n}(\vec{k})$, $\tilde{v}(\vec{k})$,
$S^{(2)}(\vec{k}_{1},\vec{k}_{2})$, and
$S^{(3)}({\vec{k}_1,\vec{k}_{2},\vec{k}_{3}})$, with $n_\mu A
\delta_{\vec{q},\vec{0}}$, $v_\mu A \delta_{\vec{q},\vec{0}}$,
$S^{(2)}_{\mu_1\mu_2}(q_1) \delta_{\vec{q}_{1}+\vec{q}_{2},\vec{0}}$, and
$S^{(3)}_{\mu_1\mu_2\mu_3}({\vec{q}_1,\vec{q}_{2},\vec{q}_{3}})$, respectively.
Eventually, one gets
\begin{multline}
   S^{(3)}_{\mu_1\mu_2\mu_3}({\vec{q}_1,\vec{q}_{2},\vec{q}_{3}})
   \approx \delta_{\vec{q}_1 + \vec{q}_2 + \vec{q}_3,\vec{0}}
   \frac{n_0^2}{L^6} \\
   \times \sum_{\substack{\beta_1,\beta_2,\beta_3 \\
       \lambda_1,\lambda_2,\lambda_3}} n_{\beta_1+\beta_2+\beta_3}
   \prod_{i=1}^3 v_{-\beta_i-\lambda_i} S^{(2)}_{\lambda_i\mu_i}(q_i).
\end{multline}

%\section{}

%\bibliographystyle{apsrev}
%\bibliographystyle{apsrev4-1}
%\bibliography{mct}

\begin{thebibliography}{72}%
\makeatletter
\providecommand \@ifxundefined [1]{%
 \@ifx{#1\undefined}
}%
\providecommand \@ifnum [1]{%
 \ifnum #1\expandafter \@firstoftwo
 \else \expandafter \@secondoftwo
 \fi
}%
\providecommand \@ifx [1]{%
 \ifx #1\expandafter \@firstoftwo
 \else \expandafter \@secondoftwo
 \fi
}%
\providecommand \natexlab [1]{#1}%
\providecommand \enquote  [1]{``#1''}%
\providecommand \bibnamefont  [1]{#1}%
\providecommand \bibfnamefont [1]{#1}%
\providecommand \citenamefont [1]{#1}%
\providecommand \href@noop [0]{\@secondoftwo}%
\providecommand \href [0]{\begingroup \@sanitize@url \@href}%
\providecommand \@href[1]{\@@startlink{#1}\@@href}%
\providecommand \@@href[1]{\endgroup#1\@@endlink}%
\providecommand \@sanitize@url [0]{\catcode `\\12\catcode `\$12\catcode
  `\&12\catcode `\#12\catcode `\^12\catcode `\_12\catcode `\%12\relax}%
\providecommand \@@startlink[1]{}%
\providecommand \@@endlink[0]{}%
\providecommand \url  [0]{\begingroup\@sanitize@url \@url }%
\providecommand \@url [1]{\endgroup\@href {#1}{\urlprefix }}%
\providecommand \urlprefix  [0]{URL }%
\providecommand \Eprint [0]{\href }%
\providecommand \doibase [0]{http://dx.doi.org/}%
\providecommand \selectlanguage [0]{\@gobble}%
\providecommand \bibinfo  [0]{\@secondoftwo}%
\providecommand \bibfield  [0]{\@secondoftwo}%
\providecommand \translation [1]{[#1]}%
\providecommand \BibitemOpen [0]{}%
\providecommand \bibitemStop [0]{}%
\providecommand \bibitemNoStop [0]{.\EOS\space}%
\providecommand \EOS [0]{\spacefactor3000\relax}%
\providecommand \BibitemShut  [1]{\csname bibitem#1\endcsname}%
\let\auto@bib@innerbib\@empty
%</preamble>
\bibitem [{\citenamefont {Bengtzelius}\ \emph {et~al.}(1984)\citenamefont
  {Bengtzelius}, \citenamefont {G\"otze},\ and\ \citenamefont
  {Sjolander}}]{Bengtzelius:1984}%
  \BibitemOpen
  \bibfield  {author} {\bibinfo {author} {\bibfnamefont {U.}~\bibnamefont
  {Bengtzelius}}, \bibinfo {author} {\bibfnamefont {W.}~\bibnamefont
  {G\"otze}}, \ and\ \bibinfo {author} {\bibfnamefont {A.}~\bibnamefont
  {Sjolander}},\ }\href {http://stacks.iop.org/0022-3719/17/i=33/a=005}
  {\bibfield  {journal} {\bibinfo  {journal} {J. Phys. C: Solid State Phys.}\
  }\textbf {\bibinfo {volume} {17}},\ \bibinfo {pages} {5915} (\bibinfo {year}
  {1984})}\BibitemShut {NoStop}%
\bibitem [{\citenamefont {G\"otze}(2009)}]{Goetze:Complex_Dynamics}%
  \BibitemOpen
  \bibfield  {author} {\bibinfo {author} {\bibfnamefont {W.}~\bibnamefont
  {G\"otze}},\ }\href@noop {} {\emph {\bibinfo {title} {Complex Dynamics of
  Glass-Forming Liquids--A Mode-Coupling Theory}}}\ (\bibinfo  {publisher}
  {Oxford},\ \bibinfo {address} {Oxford},\ \bibinfo {year} {2009})\BibitemShut
  {NoStop}%
\bibitem [{\citenamefont {G\"otze}(1999)}]{Goetze:1999}%
  \BibitemOpen
  \bibfield  {author} {\bibinfo {author} {\bibfnamefont {W.}~\bibnamefont
  {G\"otze}},\ }\href {http://stacks.iop.org/0953-8984/11/i=10A/a=002}
  {\bibfield  {journal} {\bibinfo  {journal} {J. Phys.: Condens. Matter}\
  }\textbf {\bibinfo {volume} {11}},\ \bibinfo {pages} {A1} (\bibinfo {year}
  {1999})}\BibitemShut {NoStop}%
\bibitem [{\citenamefont {Du}\ \emph {et~al.}(1994)\citenamefont {Du},
  \citenamefont {Li}, \citenamefont {Cummins}, \citenamefont {Fuchs},
  \citenamefont {Toulouse},\ and\ \citenamefont {Knauss}}]{Du:1994}%
  \BibitemOpen
  \bibfield  {author} {\bibinfo {author} {\bibfnamefont {W.~M.}\ \bibnamefont
  {Du}}, \bibinfo {author} {\bibfnamefont {G.}~\bibnamefont {Li}}, \bibinfo
  {author} {\bibfnamefont {H.~Z.}\ \bibnamefont {Cummins}}, \bibinfo {author}
  {\bibfnamefont {M.}~\bibnamefont {Fuchs}}, \bibinfo {author} {\bibfnamefont
  {J.}~\bibnamefont {Toulouse}}, \ and\ \bibinfo {author} {\bibfnamefont
  {L.~A.}\ \bibnamefont {Knauss}},\ }\href {\doibase 10.1103/PhysRevE.49.2192}
  {\bibfield  {journal} {\bibinfo  {journal} {Phys. Rev. E}\ }\textbf {\bibinfo
  {volume} {49}},\ \bibinfo {pages} {2192} (\bibinfo {year}
  {1994})}\BibitemShut {NoStop}%
\bibitem [{\citenamefont {Franosch}\ \emph
  {et~al.}(1997{\natexlab{a}})\citenamefont {Franosch}, \citenamefont
  {G\"otze}, \citenamefont {Mayr},\ and\ \citenamefont
  {Singh}}]{Franosch:1997b}%
  \BibitemOpen
  \bibfield  {author} {\bibinfo {author} {\bibfnamefont {T.}~\bibnamefont
  {Franosch}}, \bibinfo {author} {\bibfnamefont {W.}~\bibnamefont {G\"otze}},
  \bibinfo {author} {\bibfnamefont {M.~R.}\ \bibnamefont {Mayr}}, \ and\
  \bibinfo {author} {\bibfnamefont {A.~P.}\ \bibnamefont {Singh}},\ }\href
  {\doibase 10.1103/PhysRevE.55.3183} {\bibfield  {journal} {\bibinfo
  {journal} {Phys. Rev. E}\ }\textbf {\bibinfo {volume} {55}},\ \bibinfo
  {pages} {3183} (\bibinfo {year} {1997}{\natexlab{a}})}\BibitemShut {NoStop}%
\bibitem [{\citenamefont {Singh}\ \emph {et~al.}(1998)\citenamefont {Singh},
  \citenamefont {Li}, \citenamefont {G\"otze}, \citenamefont {Fuchs},
  \citenamefont {Franosch},\ and\ \citenamefont {Cummins}}]{Singh:1998}%
  \BibitemOpen
  \bibfield  {author} {\bibinfo {author} {\bibfnamefont {A.~P.}\ \bibnamefont
  {Singh}}, \bibinfo {author} {\bibfnamefont {G.}~\bibnamefont {Li}}, \bibinfo
  {author} {\bibfnamefont {W.}~\bibnamefont {G\"otze}}, \bibinfo {author}
  {\bibfnamefont {M.}~\bibnamefont {Fuchs}}, \bibinfo {author} {\bibfnamefont
  {T.}~\bibnamefont {Franosch}}, \ and\ \bibinfo {author} {\bibfnamefont
  {H.~Z.}\ \bibnamefont {Cummins}},\ }\href {\doibase
  10.1016/S0022-3093(98)00583-3} {\bibfield  {journal} {\bibinfo  {journal} {J.
  Non-Cryst. Solids}\ }\textbf {\bibinfo {volume} {235-237}},\ \bibinfo {pages}
  {66 } (\bibinfo {year} {1998})}\BibitemShut {NoStop}%
\bibitem [{\citenamefont {van Megen}\ and\ \citenamefont
  {Underwood}(1993)}]{Megen:1993}%
  \BibitemOpen
  \bibfield  {author} {\bibinfo {author} {\bibfnamefont {W.}~\bibnamefont {van
  Megen}}\ and\ \bibinfo {author} {\bibfnamefont {S.~M.}\ \bibnamefont
  {Underwood}},\ }\href {\doibase 10.1103/PhysRevE.47.248} {\bibfield
  {journal} {\bibinfo  {journal} {Phys. Rev. E}\ }\textbf {\bibinfo {volume}
  {47}},\ \bibinfo {pages} {248} (\bibinfo {year} {1993})}\BibitemShut
  {NoStop}%
\bibitem [{\citenamefont {van Megen}\ \emph {et~al.}(1998)\citenamefont {van
  Megen}, \citenamefont {Mortensen}, \citenamefont {Williams},\ and\
  \citenamefont {M\"uller}}]{Megen:1998}%
  \BibitemOpen
  \bibfield  {author} {\bibinfo {author} {\bibfnamefont {W.}~\bibnamefont {van
  Megen}}, \bibinfo {author} {\bibfnamefont {T.~C.}\ \bibnamefont {Mortensen}},
  \bibinfo {author} {\bibfnamefont {S.~R.}\ \bibnamefont {Williams}}, \ and\
  \bibinfo {author} {\bibfnamefont {J.}~\bibnamefont {M\"uller}},\ }\href
  {\doibase 10.1103/PhysRevE.58.6073} {\bibfield  {journal} {\bibinfo
  {journal} {Phys. Rev. E}\ }\textbf {\bibinfo {volume} {58}},\ \bibinfo
  {pages} {6073} (\bibinfo {year} {1998})}\BibitemShut {NoStop}%
\bibitem [{\citenamefont {Kob}\ and\ \citenamefont
  {Andersen}(1994)}]{Kob:1994}%
  \BibitemOpen
  \bibfield  {author} {\bibinfo {author} {\bibfnamefont {W.}~\bibnamefont
  {Kob}}\ and\ \bibinfo {author} {\bibfnamefont {H.~C.}\ \bibnamefont
  {Andersen}},\ }\href {\doibase 10.1103/PhysRevLett.73.1376} {\bibfield
  {journal} {\bibinfo  {journal} {Phys. Rev. Lett.}\ }\textbf {\bibinfo
  {volume} {73}},\ \bibinfo {pages} {1376} (\bibinfo {year}
  {1994})}\BibitemShut {NoStop}%
\bibitem [{\citenamefont {Kob}\ and\ \citenamefont
  {Andersen}(1995)}]{Kob:1995}%
  \BibitemOpen
  \bibfield  {author} {\bibinfo {author} {\bibfnamefont {W.}~\bibnamefont
  {Kob}}\ and\ \bibinfo {author} {\bibfnamefont {H.~C.}\ \bibnamefont
  {Andersen}},\ }\href {\doibase 10.1103/PhysRevE.51.4626} {\bibfield
  {journal} {\bibinfo  {journal} {Phys. Rev. E}\ }\textbf {\bibinfo {volume}
  {51}},\ \bibinfo {pages} {4626} (\bibinfo {year} {1995})}\BibitemShut
  {NoStop}%
\bibitem [{\citenamefont {Bayer}\ \emph {et~al.}(2007)\citenamefont {Bayer},
  \citenamefont {Brader}, \citenamefont {Ebert}, \citenamefont {Fuchs},
  \citenamefont {Lange}, \citenamefont {Maret}, \citenamefont {Schilling},
  \citenamefont {Sperl},\ and\ \citenamefont {Wittmer}}]{Bayer:2007}%
  \BibitemOpen
  \bibfield  {author} {\bibinfo {author} {\bibfnamefont {M.}~\bibnamefont
  {Bayer}}, \bibinfo {author} {\bibfnamefont {J.~M.}\ \bibnamefont {Brader}},
  \bibinfo {author} {\bibfnamefont {F.}~\bibnamefont {Ebert}}, \bibinfo
  {author} {\bibfnamefont {M.}~\bibnamefont {Fuchs}}, \bibinfo {author}
  {\bibfnamefont {E.}~\bibnamefont {Lange}}, \bibinfo {author} {\bibfnamefont
  {G.}~\bibnamefont {Maret}}, \bibinfo {author} {\bibfnamefont
  {R.}~\bibnamefont {Schilling}}, \bibinfo {author} {\bibfnamefont
  {M.}~\bibnamefont {Sperl}}, \ and\ \bibinfo {author} {\bibfnamefont {J.~P.}\
  \bibnamefont {Wittmer}},\ }\href {\doibase 10.1103/PhysRevE.76.011508}
  {\bibfield  {journal} {\bibinfo  {journal} {Phys. Rev. E}\ }\textbf {\bibinfo
  {volume} {76}},\ \bibinfo {pages} {011508} (\bibinfo {year}
  {2007})}\BibitemShut {NoStop}%
\bibitem [{\citenamefont {Hajnal}\ \emph {et~al.}(2009)\citenamefont {Hajnal},
  \citenamefont {Brader},\ and\ \citenamefont {Schilling}}]{Hajnal:2009}%
  \BibitemOpen
  \bibfield  {author} {\bibinfo {author} {\bibfnamefont {D.}~\bibnamefont
  {Hajnal}}, \bibinfo {author} {\bibfnamefont {J.~M.}\ \bibnamefont {Brader}},
  \ and\ \bibinfo {author} {\bibfnamefont {R.}~\bibnamefont {Schilling}},\
  }\href {\doibase 10.1103/PhysRevE.80.021503} {\bibfield  {journal} {\bibinfo
  {journal} {Phys. Rev. E}\ }\textbf {\bibinfo {volume} {80}},\ \bibinfo
  {pages} {021503} (\bibinfo {year} {2009})}\BibitemShut {NoStop}%
\bibitem [{\citenamefont {Hajnal}\ \emph {et~al.}(2011)\citenamefont {Hajnal},
  \citenamefont {Oettel},\ and\ \citenamefont {Schilling}}]{Hajnal:2011}%
  \BibitemOpen
  \bibfield  {author} {\bibinfo {author} {\bibfnamefont {D.}~\bibnamefont
  {Hajnal}}, \bibinfo {author} {\bibfnamefont {M.}~\bibnamefont {Oettel}}, \
  and\ \bibinfo {author} {\bibfnamefont {R.}~\bibnamefont {Schilling}},\ }\href
  {\doibase 10.1016/j.jnoncrysol.2010.06.039} {\bibfield  {journal} {\bibinfo
  {journal} {J. Non-Cryst. Solids}\ }\textbf {\bibinfo {volume} {357}},\
  \bibinfo {pages} {302 } (\bibinfo {year} {2011})}\BibitemShut {NoStop}%
\bibitem [{\citenamefont {Schmid}\ and\ \citenamefont
  {Schilling}(2010)}]{Schmid:2010}%
  \BibitemOpen
  \bibfield  {author} {\bibinfo {author} {\bibfnamefont {B.}~\bibnamefont
  {Schmid}}\ and\ \bibinfo {author} {\bibfnamefont {R.}~\bibnamefont
  {Schilling}},\ }\href {\doibase 10.1103/PhysRevE.81.041502} {\bibfield
  {journal} {\bibinfo  {journal} {Phys. Rev. E}\ }\textbf {\bibinfo {volume}
  {81}},\ \bibinfo {pages} {041502} (\bibinfo {year} {2010})}\BibitemShut
  {NoStop}%
\bibitem [{\citenamefont {Ikeda}\ and\ \citenamefont
  {Miyazaki}(2010)}]{Miyazaki:2010}%
  \BibitemOpen
  \bibfield  {author} {\bibinfo {author} {\bibfnamefont {A.}~\bibnamefont
  {Ikeda}}\ and\ \bibinfo {author} {\bibfnamefont {K.}~\bibnamefont
  {Miyazaki}},\ }\href {\doibase 10.1103/PhysRevLett.104.255704} {\bibfield
  {journal} {\bibinfo  {journal} {Phys. Rev. Lett.}\ }\textbf {\bibinfo
  {volume} {104}},\ \bibinfo {pages} {255704} (\bibinfo {year}
  {2010})}\BibitemShut {NoStop}%
\bibitem [{\citenamefont {Schilling}\ and\ \citenamefont
  {Schmid}(2011)}]{Schmid:2011}%
  \BibitemOpen
  \bibfield  {author} {\bibinfo {author} {\bibfnamefont {R.}~\bibnamefont
  {Schilling}}\ and\ \bibinfo {author} {\bibfnamefont {B.}~\bibnamefont
  {Schmid}},\ }\href {\doibase 10.1103/PhysRevLett.106.049601} {\bibfield
  {journal} {\bibinfo  {journal} {Phys. Rev. Lett.}\ }\textbf {\bibinfo
  {volume} {106}},\ \bibinfo {pages} {049601} (\bibinfo {year}
  {2011})}\BibitemShut {NoStop}%
\bibitem [{\citenamefont {Ikeda}\ and\ \citenamefont
  {Miyazaki}(2011)}]{Miyazaki:2011}%
  \BibitemOpen
  \bibfield  {author} {\bibinfo {author} {\bibfnamefont {A.}~\bibnamefont
  {Ikeda}}\ and\ \bibinfo {author} {\bibfnamefont {K.}~\bibnamefont
  {Miyazaki}},\ }\href {\doibase 10.1103/PhysRevLett.106.049602} {\bibfield
  {journal} {\bibinfo  {journal} {Phys. Rev. Lett.}\ }\textbf {\bibinfo
  {volume} {106}},\ \bibinfo {pages} {049602} (\bibinfo {year}
  {2011})}\BibitemShut {NoStop}%
\bibitem [{\citenamefont {Schilling}\ and\ \citenamefont
  {Scheidsteger}(1997)}]{Scheidsteger:1997}%
  \BibitemOpen
  \bibfield  {author} {\bibinfo {author} {\bibfnamefont {R.}~\bibnamefont
  {Schilling}}\ and\ \bibinfo {author} {\bibfnamefont {T.}~\bibnamefont
  {Scheidsteger}},\ }\href {\doibase 10.1103/PhysRevE.56.2932} {\bibfield
  {journal} {\bibinfo  {journal} {Phys. Rev. E}\ }\textbf {\bibinfo {volume}
  {56}},\ \bibinfo {pages} {2932} (\bibinfo {year} {1997})}\BibitemShut
  {NoStop}%
\bibitem [{\citenamefont {Franosch}\ \emph
  {et~al.}(1997{\natexlab{b}})\citenamefont {Franosch}, \citenamefont {Fuchs},
  \citenamefont {G{\"o}tze}, \citenamefont {Mayr},\ and\ \citenamefont
  {Singh}}]{Franosch:1997}%
  \BibitemOpen
  \bibfield  {author} {\bibinfo {author} {\bibfnamefont {T.}~\bibnamefont
  {Franosch}}, \bibinfo {author} {\bibfnamefont {M.}~\bibnamefont {Fuchs}},
  \bibinfo {author} {\bibfnamefont {W.}~\bibnamefont {G{\"o}tze}}, \bibinfo
  {author} {\bibfnamefont {M.~R.}\ \bibnamefont {Mayr}}, \ and\ \bibinfo
  {author} {\bibfnamefont {A.~P.}\ \bibnamefont {Singh}},\ }\href {\doibase
  10.1103/PhysRevE.56.5659} {\bibfield  {journal} {\bibinfo  {journal} {{Phys.
  Rev. E}}\ }\textbf {\bibinfo {volume} {{56}}},\ \bibinfo {pages} {5659}
  (\bibinfo {year} {1997}{\natexlab{b}})}\BibitemShut {NoStop}%
\bibitem [{\citenamefont {Scheidsteger}\ and\ \citenamefont
  {Schilling}(1998)}]{Scheidsteger:1998}%
  \BibitemOpen
  \bibfield  {author} {\bibinfo {author} {\bibfnamefont {T.}~\bibnamefont
  {Scheidsteger}}\ and\ \bibinfo {author} {\bibfnamefont {R.}~\bibnamefont
  {Schilling}},\ }\href {\doibase 10.1080/13642819808204956} {\bibfield
  {journal} {\bibinfo  {journal} {Phil. Mag. B}\ }\textbf {\bibinfo {volume}
  {77}},\ \bibinfo {pages} {305} (\bibinfo {year} {1998})}\BibitemShut
  {NoStop}%
\bibitem [{\citenamefont {K\"ammerer}\ \emph
  {et~al.}(1998{\natexlab{a}})\citenamefont {K\"ammerer}, \citenamefont {Kob},\
  and\ \citenamefont {Schilling}}]{Kaemmerer:1998}%
  \BibitemOpen
  \bibfield  {author} {\bibinfo {author} {\bibfnamefont {S.}~\bibnamefont
  {K\"ammerer}}, \bibinfo {author} {\bibfnamefont {W.}~\bibnamefont {Kob}}, \
  and\ \bibinfo {author} {\bibfnamefont {R.}~\bibnamefont {Schilling}},\ }\href
  {\doibase 10.1103/PhysRevE.58.2131} {\bibfield  {journal} {\bibinfo
  {journal} {Phys. Rev. E}\ }\textbf {\bibinfo {volume} {58}},\ \bibinfo
  {pages} {2131} (\bibinfo {year} {1998}{\natexlab{a}})}\BibitemShut {NoStop}%
\bibitem [{\citenamefont {K\"ammerer}\ \emph
  {et~al.}(1998{\natexlab{b}})\citenamefont {K\"ammerer}, \citenamefont {Kob},\
  and\ \citenamefont {Schilling}}]{Kaemmerer:1998a}%
  \BibitemOpen
  \bibfield  {author} {\bibinfo {author} {\bibfnamefont {S.}~\bibnamefont
  {K\"ammerer}}, \bibinfo {author} {\bibfnamefont {W.}~\bibnamefont {Kob}}, \
  and\ \bibinfo {author} {\bibfnamefont {R.}~\bibnamefont {Schilling}},\ }\href
  {\doibase 10.1103/PhysRevE.58.2141} {\bibfield  {journal} {\bibinfo
  {journal} {Phys. Rev. E}\ }\textbf {\bibinfo {volume} {58}},\ \bibinfo
  {pages} {2141} (\bibinfo {year} {1998}{\natexlab{b}})}\BibitemShut {NoStop}%
\bibitem [{\citenamefont {Fabbian}\ \emph {et~al.}(2000)\citenamefont
  {Fabbian}, \citenamefont {Latz}, \citenamefont {Schilling}, \citenamefont
  {Sciortino}, \citenamefont {Tartaglia},\ and\ \citenamefont
  {Theis}}]{Fabbian:1999}%
  \BibitemOpen
  \bibfield  {author} {\bibinfo {author} {\bibfnamefont {L.}~\bibnamefont
  {Fabbian}}, \bibinfo {author} {\bibfnamefont {A.}~\bibnamefont {Latz}},
  \bibinfo {author} {\bibfnamefont {R.}~\bibnamefont {Schilling}}, \bibinfo
  {author} {\bibfnamefont {F.}~\bibnamefont {Sciortino}}, \bibinfo {author}
  {\bibfnamefont {P.}~\bibnamefont {Tartaglia}}, \ and\ \bibinfo {author}
  {\bibfnamefont {C.}~\bibnamefont {Theis}},\ }\href {\doibase
  10.1103/PhysRevE.62.2388} {\bibfield  {journal} {\bibinfo  {journal} {Phys.
  Rev. E}\ }\textbf {\bibinfo {volume} {62}},\ \bibinfo {pages} {2388}
  (\bibinfo {year} {2000})}\BibitemShut {NoStop}%
\bibitem [{\citenamefont {Schilling}(2002)}]{Schilling:2002}%
  \BibitemOpen
  \bibfield  {author} {\bibinfo {author} {\bibfnamefont {R.}~\bibnamefont
  {Schilling}},\ }\href {\doibase 10.1103/PhysRevE.65.051206} {\bibfield
  {journal} {\bibinfo  {journal} {Phys. Rev. E}\ }\textbf {\bibinfo {volume}
  {65}},\ \bibinfo {pages} {051206} (\bibinfo {year} {2002})}\BibitemShut
  {NoStop}%
\bibitem [{\citenamefont {Nandi}\ \emph {et~al.}(2011)\citenamefont {Nandi},
  \citenamefont {Bhattacharyya},\ and\ \citenamefont
  {Ramaswamy}}]{Ramaswamy:2011}%
  \BibitemOpen
  \bibfield  {author} {\bibinfo {author} {\bibfnamefont {S.~K.}\ \bibnamefont
  {Nandi}}, \bibinfo {author} {\bibfnamefont {S.~M.}\ \bibnamefont
  {Bhattacharyya}}, \ and\ \bibinfo {author} {\bibfnamefont {S.}~\bibnamefont
  {Ramaswamy}},\ }\href {\doibase 10.1103/PhysRevE.84.061501} {\bibfield
  {journal} {\bibinfo  {journal} {Phys. Rev. E}\ }\textbf {\bibinfo {volume}
  {84}},\ \bibinfo {pages} {061501} (\bibinfo {year} {2011})}\BibitemShut
  {NoStop}%
\bibitem [{\citenamefont {Krakoviack}(2005)}]{Krakoviack:2005}%
  \BibitemOpen
  \bibfield  {author} {\bibinfo {author} {\bibfnamefont {V.}~\bibnamefont
  {Krakoviack}},\ }\href {\doibase 10.1103/PhysRevLett.94.065703} {\bibfield
  {journal} {\bibinfo  {journal} {Phys. Rev. Lett.}\ }\textbf {\bibinfo
  {volume} {94}},\ \bibinfo {pages} {065703} (\bibinfo {year}
  {2005})}\BibitemShut {NoStop}%
\bibitem [{\citenamefont {Krakoviack}(2007)}]{Krakoviack:2007}%
  \BibitemOpen
  \bibfield  {author} {\bibinfo {author} {\bibfnamefont {V.}~\bibnamefont
  {Krakoviack}},\ }\href {\doibase 10.1103/PhysRevE.75.031503} {\bibfield
  {journal} {\bibinfo  {journal} {Phys. Rev. E}\ }\textbf {\bibinfo {volume}
  {75}},\ \bibinfo {eid} {031503} (\bibinfo {year} {2007})}\BibitemShut
  {NoStop}%
\bibitem [{\citenamefont {Krakoviack}(2009)}]{Krakoviack:2009}%
  \BibitemOpen
  \bibfield  {author} {\bibinfo {author} {\bibfnamefont {V.}~\bibnamefont
  {Krakoviack}},\ }\href {\doibase 10.1103/PhysRevE.79.061501} {\bibfield
  {journal} {\bibinfo  {journal} {Phys. Rev. E}\ }\textbf {\bibinfo {volume}
  {79}},\ \bibinfo {eid} {061501} (\bibinfo {year} {2009})}\BibitemShut
  {NoStop}%
\bibitem [{\citenamefont {Krakoviack}(2011)}]{Krakoviack:2011}%
  \BibitemOpen
  \bibfield  {author} {\bibinfo {author} {\bibfnamefont {V.}~\bibnamefont
  {Krakoviack}},\ }\href {\doibase 10.1103/PhysRevE.84.050501} {\bibfield
  {journal} {\bibinfo  {journal} {Phys. Rev. E}\ }\textbf {\bibinfo {volume}
  {84}},\ \bibinfo {pages} {050501} (\bibinfo {year} {2011})}\BibitemShut
  {NoStop}%
\bibitem [{\citenamefont {Schnyder}\ \emph {et~al.}(2011)\citenamefont
  {Schnyder}, \citenamefont {H\"ofling}, \citenamefont {Franosch},\ and\
  \citenamefont {$\text{Th.}$ Voigtmann}}]{Schnyder:2011}%
  \BibitemOpen
  \bibfield  {author} {\bibinfo {author} {\bibfnamefont {S.~K.}\ \bibnamefont
  {Schnyder}}, \bibinfo {author} {\bibfnamefont {F.}~\bibnamefont {H\"ofling}},
  \bibinfo {author} {\bibfnamefont {T.}~\bibnamefont {Franosch}}, \ and\
  \bibinfo {author} {\bibnamefont {$\text{Th.}$ Voigtmann}},\ }\href {\doibase
  10.1088/0953-8984/23/23/234121} {\bibfield  {journal} {\bibinfo  {journal}
  {J. Phys. Condens. Matter}\ }\textbf {\bibinfo {volume} {23}},\ \bibinfo
  {pages} {234121} (\bibinfo {year} {2011})}\BibitemShut {NoStop}%
\bibitem [{\citenamefont {H{\"o}f\/ling}\ \emph {et~al.}(2006)\citenamefont
  {H{\"o}f\/ling}, \citenamefont {Franosch},\ and\ \citenamefont
  {Frey}}]{Lorentz_PRL:2006}%
  \BibitemOpen
  \bibfield  {author} {\bibinfo {author} {\bibfnamefont {F.}~\bibnamefont
  {H{\"o}f\/ling}}, \bibinfo {author} {\bibfnamefont {T.}~\bibnamefont
  {Franosch}}, \ and\ \bibinfo {author} {\bibfnamefont {E.}~\bibnamefont
  {Frey}},\ }\href {\doibase 10.1103/PhysRevLett.96.165901} {\bibfield
  {journal} {\bibinfo  {journal} {Phys. Rev. Lett.}\ }\textbf {\bibinfo
  {volume} {96}},\ \bibinfo {pages} {165901} (\bibinfo {year}
  {2006})}\BibitemShut {NoStop}%
\bibitem [{\citenamefont {H{\"o}f\/ling}\ \emph {et~al.}(2008)\citenamefont
  {H{\"o}f\/ling}, \citenamefont {Munk}, \citenamefont {Frey},\ and\
  \citenamefont {Franosch}}]{Lorentz_JCP:2008}%
  \BibitemOpen
  \bibfield  {author} {\bibinfo {author} {\bibfnamefont {F.}~\bibnamefont
  {H{\"o}f\/ling}}, \bibinfo {author} {\bibfnamefont {T.}~\bibnamefont {Munk}},
  \bibinfo {author} {\bibfnamefont {E.}~\bibnamefont {Frey}}, \ and\ \bibinfo
  {author} {\bibfnamefont {T.}~\bibnamefont {Franosch}},\ }\href {\doibase
  10.1063/1.2901170} {\bibfield  {journal} {\bibinfo  {journal} {J. Chem.
  Phys.}\ }\textbf {\bibinfo {volume} {128}},\ \bibinfo {eid} {164517}
  (\bibinfo {year} {2008})}\BibitemShut {NoStop}%
\bibitem [{\citenamefont {Franosch}\ \emph {et~al.}(2011)\citenamefont
  {Franosch}, \citenamefont {Spanner}, \citenamefont {Bauer}, \citenamefont
  {Schr{\"o}der-Turk},\ and\ \citenamefont {H{\"o}fling}}]{Franosch:2011}%
  \BibitemOpen
  \bibfield  {author} {\bibinfo {author} {\bibfnamefont {T.}~\bibnamefont
  {Franosch}}, \bibinfo {author} {\bibfnamefont {M.}~\bibnamefont {Spanner}},
  \bibinfo {author} {\bibfnamefont {T.}~\bibnamefont {Bauer}}, \bibinfo
  {author} {\bibfnamefont {G.~E.}\ \bibnamefont {Schr{\"o}der-Turk}}, \ and\
  \bibinfo {author} {\bibfnamefont {F.}~\bibnamefont {H{\"o}fling}},\ }\href
  {\doibase {10.1016/j.jnoncrysol.2010.06.051}} {\bibfield  {journal} {\bibinfo
   {journal} {{J. Non-Cryst. Solids}}\ }\textbf {\bibinfo {volume} {{357}}},\
  \bibinfo {pages} {472} (\bibinfo {year} {{2011}})}\BibitemShut {NoStop}%
\bibitem [{\citenamefont {Biroli}\ \emph {et~al.}(2006)\citenamefont {Biroli},
  \citenamefont {Bouchaud}, \citenamefont {Miyazaki},\ and\ \citenamefont
  {Reichman}}]{Biroli:2006}%
  \BibitemOpen
  \bibfield  {author} {\bibinfo {author} {\bibfnamefont {G.}~\bibnamefont
  {Biroli}}, \bibinfo {author} {\bibfnamefont {J.-P.}\ \bibnamefont
  {Bouchaud}}, \bibinfo {author} {\bibfnamefont {K.}~\bibnamefont {Miyazaki}},
  \ and\ \bibinfo {author} {\bibfnamefont {D.~R.}\ \bibnamefont {Reichman}},\
  }\href {\doibase 10.1103/PhysRevLett.97.195701} {\bibfield  {journal}
  {\bibinfo  {journal} {Phys. Rev. Lett.}\ }\textbf {\bibinfo {volume} {97}},\
  \bibinfo {pages} {195701} (\bibinfo {year} {2006})}\BibitemShut {NoStop}%
\bibitem [{Wor(2000)}]{Workshop_Confinement:2000}%
  \BibitemOpen
  \href@noop {} {\emph {\bibinfo {title} {Special Issue: International Workshop
  on Dynamics in Confinement}}},\ Vol.~\bibinfo {volume} {10}\ (\bibinfo {year}
  {2000})\ \bibinfo {note} {$\text{J}$. Phys. IV}\BibitemShut {NoStop}%
\bibitem [{\citenamefont {Scheidler}\ \emph
  {et~al.}(2000{\natexlab{a}})\citenamefont {Scheidler}, \citenamefont {Kob},\
  and\ \citenamefont {Binder}}]{Scheidler:2000}%
  \BibitemOpen
  \bibfield  {author} {\bibinfo {author} {\bibfnamefont {P.}~\bibnamefont
  {Scheidler}}, \bibinfo {author} {\bibfnamefont {W.}~\bibnamefont {Kob}}, \
  and\ \bibinfo {author} {\bibfnamefont {K.}~\bibnamefont {Binder}},\ }\href
  {\doibase 10.1051/jp4:2000706} {\bibfield  {journal} {\bibinfo  {journal} {J.
  Phys. IV France}\ }\textbf {\bibinfo {volume} {10}},\ \bibinfo {pages} {Pr7}
  (\bibinfo {year} {2000}{\natexlab{a}})}\BibitemShut {NoStop}%
\bibitem [{\citenamefont {Scheidler}\ \emph
  {et~al.}(2000{\natexlab{b}})\citenamefont {Scheidler}, \citenamefont {Kob},\
  and\ \citenamefont {Binder}}]{Scheidler:2000a}%
  \BibitemOpen
  \bibfield  {author} {\bibinfo {author} {\bibfnamefont {P.}~\bibnamefont
  {Scheidler}}, \bibinfo {author} {\bibfnamefont {W.}~\bibnamefont {Kob}}, \
  and\ \bibinfo {author} {\bibfnamefont {K.}~\bibnamefont {Binder}},\ }\href
  {http://stacks.iop.org/0295-5075/52/i=3/a=277} {\bibfield  {journal}
  {\bibinfo  {journal} {Europhys. Lett.}\ }\textbf {\bibinfo {volume} {52}},\
  \bibinfo {pages} {277} (\bibinfo {year} {2000}{\natexlab{b}})}\BibitemShut
  {NoStop}%
\bibitem [{\citenamefont {Scheidler}\ \emph {et~al.}(2004)\citenamefont
  {Scheidler}, \citenamefont {Kob},\ and\ \citenamefont
  {Binder}}]{Scheidler:2004}%
  \BibitemOpen
  \bibfield  {author} {\bibinfo {author} {\bibfnamefont {P.}~\bibnamefont
  {Scheidler}}, \bibinfo {author} {\bibfnamefont {W.}~\bibnamefont {Kob}}, \
  and\ \bibinfo {author} {\bibfnamefont {K.}~\bibnamefont {Binder}},\ }\href
  {\doibase 10.1021/jp036593s} {\bibfield  {journal} {\bibinfo  {journal} {J.
  Phys. Chem. B}\ }\textbf {\bibinfo {volume} {108}},\ \bibinfo {pages} {6673}
  (\bibinfo {year} {2004})}\BibitemShut {NoStop}%
\bibitem [{\citenamefont {Teboul}\ and\ \citenamefont {{Alba
  Simionesco}}(2002)}]{Teboul:2002}%
  \BibitemOpen
  \bibfield  {author} {\bibinfo {author} {\bibfnamefont {V.}~\bibnamefont
  {Teboul}}\ and\ \bibinfo {author} {\bibfnamefont {C.}~\bibnamefont {{Alba
  Simionesco}}},\ }\href {\doibase 10.1088/0953-8984/14/23/304} {\bibfield
  {journal} {\bibinfo  {journal} {J. Phys. Condens. Matter}\ }\textbf {\bibinfo
  {volume} {14}},\ \bibinfo {pages} {5699} (\bibinfo {year}
  {2002})}\BibitemShut {NoStop}%
\bibitem [{\citenamefont {Fehr}\ and\ \citenamefont
  {L\"owen}(1995)}]{Fehr:1995}%
  \BibitemOpen
  \bibfield  {author} {\bibinfo {author} {\bibfnamefont {T.}~\bibnamefont
  {Fehr}}\ and\ \bibinfo {author} {\bibfnamefont {H.}~\bibnamefont {L\"owen}},\
  }\href {\doibase 10.1103/PhysRevE.52.4016} {\bibfield  {journal} {\bibinfo
  {journal} {Phys. Rev. E}\ }\textbf {\bibinfo {volume} {52}},\ \bibinfo
  {pages} {4016} (\bibinfo {year} {1995})}\BibitemShut {NoStop}%
\bibitem [{\citenamefont {Varnik}\ \emph {et~al.}(2000)\citenamefont {Varnik},
  \citenamefont {Baschnagel},\ and\ \citenamefont {Binder}}]{Varnik:2000}%
  \BibitemOpen
  \bibfield  {author} {\bibinfo {author} {\bibfnamefont {F.}~\bibnamefont
  {Varnik}}, \bibinfo {author} {\bibfnamefont {J.}~\bibnamefont {Baschnagel}},
  \ and\ \bibinfo {author} {\bibfnamefont {K.}~\bibnamefont {Binder}},\ }\href
  {\doibase 10.1051/jp4:2000747} {\bibfield  {journal} {\bibinfo  {journal} {J.
  Phys. IV}\ }\textbf {\bibinfo {volume} {10}},\ \bibinfo {pages} {239}
  (\bibinfo {year} {2000})}\BibitemShut {NoStop}%
\bibitem [{\citenamefont {Varnik}\ \emph {et~al.}(2002)\citenamefont {Varnik},
  \citenamefont {Baschnagel},\ and\ \citenamefont {Binder}}]{Varnik:2002}%
  \BibitemOpen
  \bibfield  {author} {\bibinfo {author} {\bibfnamefont {F.}~\bibnamefont
  {Varnik}}, \bibinfo {author} {\bibfnamefont {J.}~\bibnamefont {Baschnagel}},
  \ and\ \bibinfo {author} {\bibfnamefont {K.}~\bibnamefont {Binder}},\ }\href
  {\doibase 10.1103/PhysRevE.65.021507} {\bibfield  {journal} {\bibinfo
  {journal} {Phys. Rev. E}\ }\textbf {\bibinfo {volume} {65}},\ \bibinfo
  {pages} {021507} (\bibinfo {year} {2002})}\BibitemShut {NoStop}%
\bibitem [{\citenamefont {Baschnagel}\ and\ \citenamefont
  {Varnik}(2005)}]{Baschnagel:2005}%
  \BibitemOpen
  \bibfield  {author} {\bibinfo {author} {\bibfnamefont {J.}~\bibnamefont
  {Baschnagel}}\ and\ \bibinfo {author} {\bibfnamefont {F.}~\bibnamefont
  {Varnik}},\ }\href {\doibase 10.1088/0953-8984/17/32/R02} {\bibfield
  {journal} {\bibinfo  {journal} {J. Phys. Condens. Matter}\ }\textbf {\bibinfo
  {volume} {17}},\ \bibinfo {pages} {R851} (\bibinfo {year}
  {2005})}\BibitemShut {NoStop}%
\bibitem [{\citenamefont {Nugent}\ \emph {et~al.}(2007)\citenamefont {Nugent},
  \citenamefont {Edmond}, \citenamefont {Patel},\ and\ \citenamefont
  {Weeks}}]{Nugent:2007}%
  \BibitemOpen
  \bibfield  {author} {\bibinfo {author} {\bibfnamefont {C.~R.}\ \bibnamefont
  {Nugent}}, \bibinfo {author} {\bibfnamefont {K.~V.}\ \bibnamefont {Edmond}},
  \bibinfo {author} {\bibfnamefont {H.~N.}\ \bibnamefont {Patel}}, \ and\
  \bibinfo {author} {\bibfnamefont {E.~R.}\ \bibnamefont {Weeks}},\ }\href
  {\doibase 10.1103/PhysRevLett.99.025702} {\bibfield  {journal} {\bibinfo
  {journal} {Phys. Rev. Lett.}\ }\textbf {\bibinfo {volume} {99}},\ \bibinfo
  {pages} {025702} (\bibinfo {year} {2007})}\BibitemShut {NoStop}%
\bibitem [{\citenamefont {Eral}\ \emph {et~al.}(2009)\citenamefont {Eral},
  \citenamefont {van~den Ende}, \citenamefont {Mugele},\ and\ \citenamefont
  {Duits}}]{Eral:2009}%
  \BibitemOpen
  \bibfield  {author} {\bibinfo {author} {\bibfnamefont {H.~B.}\ \bibnamefont
  {Eral}}, \bibinfo {author} {\bibfnamefont {D.}~\bibnamefont {van~den Ende}},
  \bibinfo {author} {\bibfnamefont {F.}~\bibnamefont {Mugele}}, \ and\ \bibinfo
  {author} {\bibfnamefont {M.~H.~G.}\ \bibnamefont {Duits}},\ }\href {\doibase
  10.1103/PhysRevE.80.061403} {\bibfield  {journal} {\bibinfo  {journal} {Phys.
  Rev. E}\ }\textbf {\bibinfo {volume} {80}},\ \bibinfo {pages} {061403}
  (\bibinfo {year} {2009})}\BibitemShut {NoStop}%
\bibitem [{\citenamefont {Eral}\ \emph {et~al.}(2011)\citenamefont {Eral},
  \citenamefont {Mugele},\ and\ \citenamefont {Duits}}]{Eral:2011}%
  \BibitemOpen
  \bibfield  {author} {\bibinfo {author} {\bibfnamefont {H.~B.}\ \bibnamefont
  {Eral}}, \bibinfo {author} {\bibfnamefont {F.}~\bibnamefont {Mugele}}, \ and\
  \bibinfo {author} {\bibfnamefont {M.~H.~G.}\ \bibnamefont {Duits}},\ }\href
  {\doibase 10.1021/la2024764} {\bibfield  {journal} {\bibinfo  {journal}
  {Langmuir}\ }\textbf {\bibinfo {volume} {27}},\ \bibinfo {pages} {12297}
  (\bibinfo {year} {2011})}\BibitemShut {NoStop}%
\bibitem [{\citenamefont {Edmond}\ \emph {et~al.}(2010)\citenamefont {Edmond},
  \citenamefont {Nugent},\ and\ \citenamefont {Weeks}}]{Edmond:2010}%
  \BibitemOpen
  \bibfield  {author} {\bibinfo {author} {\bibfnamefont {K.}~\bibnamefont
  {Edmond}}, \bibinfo {author} {\bibfnamefont {C.}~\bibnamefont {Nugent}}, \
  and\ \bibinfo {author} {\bibfnamefont {E.}~\bibnamefont {Weeks}},\ }\href
  {http://dx.doi.org/10.1140/epjst/e2010-01311-3} {\bibfield  {journal}
  {\bibinfo  {journal} {Eur. Phys. J. Spec. Top.}\ }\textbf {\bibinfo {volume}
  {189}},\ \bibinfo {pages} {83} (\bibinfo {year} {2010})}\BibitemShut
  {NoStop}%
\bibitem [{\citenamefont {Mittal}\ \emph {et~al.}(2008)\citenamefont {Mittal},
  \citenamefont {Truskett}, \citenamefont {Errington},\ and\ \citenamefont
  {Hummer}}]{Mittal:2008}%
  \BibitemOpen
  \bibfield  {author} {\bibinfo {author} {\bibfnamefont {J.}~\bibnamefont
  {Mittal}}, \bibinfo {author} {\bibfnamefont {T.~M.}\ \bibnamefont
  {Truskett}}, \bibinfo {author} {\bibfnamefont {J.~R.}\ \bibnamefont
  {Errington}}, \ and\ \bibinfo {author} {\bibfnamefont {G.}~\bibnamefont
  {Hummer}},\ }\href {\doibase 10.1103/PhysRevLett.100.145901} {\bibfield
  {journal} {\bibinfo  {journal} {Phys. Rev. Lett.}\ }\textbf {\bibinfo
  {volume} {100}},\ \bibinfo {pages} {145901} (\bibinfo {year}
  {2008})}\BibitemShut {NoStop}%
\bibitem [{\citenamefont {Goel}\ \emph {et~al.}(2009)\citenamefont {Goel},
  \citenamefont {Krekelberg}, \citenamefont {Pond}, \citenamefont {Mittal},
  \citenamefont {Shen}, \citenamefont {Errington},\ and\ \citenamefont
  {Truskett}}]{Goel:2009}%
  \BibitemOpen
  \bibfield  {author} {\bibinfo {author} {\bibfnamefont {G.}~\bibnamefont
  {Goel}}, \bibinfo {author} {\bibfnamefont {W.~P.}\ \bibnamefont
  {Krekelberg}}, \bibinfo {author} {\bibfnamefont {M.~J.}\ \bibnamefont
  {Pond}}, \bibinfo {author} {\bibfnamefont {J.}~\bibnamefont {Mittal}},
  \bibinfo {author} {\bibfnamefont {V.~K.}\ \bibnamefont {Shen}}, \bibinfo
  {author} {\bibfnamefont {J.~R.}\ \bibnamefont {Errington}}, \ and\ \bibinfo
  {author} {\bibfnamefont {T.~M.}\ \bibnamefont {Truskett}},\ }\href
  {http://stacks.iop.org/1742-5468/2009/i=04/a=P04006} {\bibfield  {journal}
  {\bibinfo  {journal} {J. Stat. Mech.}\ }\textbf {\bibinfo {volume} {2009}},\
  \bibinfo {pages} {P04006} (\bibinfo {year} {2009})}\BibitemShut {NoStop}%
\bibitem [{\citenamefont {Gallo}\ \emph
  {et~al.}(2000{\natexlab{a}})\citenamefont {Gallo}, \citenamefont {Rovere},\
  and\ \citenamefont {Spohr}}]{Gallo:2000a}%
  \BibitemOpen
  \bibfield  {author} {\bibinfo {author} {\bibfnamefont {P.}~\bibnamefont
  {Gallo}}, \bibinfo {author} {\bibfnamefont {M.}~\bibnamefont {Rovere}}, \
  and\ \bibinfo {author} {\bibfnamefont {E.}~\bibnamefont {Spohr}},\ }\href
  {\doibase 10.1103/PhysRevLett.85.4317} {\bibfield  {journal} {\bibinfo
  {journal} {Phys. Rev. Lett.}\ }\textbf {\bibinfo {volume} {85}},\ \bibinfo
  {pages} {4317} (\bibinfo {year} {2000}{\natexlab{a}})}\BibitemShut {NoStop}%
\bibitem [{\citenamefont {Gallo}\ \emph
  {et~al.}(2000{\natexlab{b}})\citenamefont {Gallo}, \citenamefont {Rovere},\
  and\ \citenamefont {Spohr}}]{Gallo:2000b}%
  \BibitemOpen
  \bibfield  {author} {\bibinfo {author} {\bibfnamefont {P.}~\bibnamefont
  {Gallo}}, \bibinfo {author} {\bibfnamefont {M.}~\bibnamefont {Rovere}}, \
  and\ \bibinfo {author} {\bibfnamefont {E.}~\bibnamefont {Spohr}},\ }\href
  {\doibase 10.1063/1.1328073} {\bibfield  {journal} {\bibinfo  {journal} {J.
  Chem. Phys.}\ }\textbf {\bibinfo {volume} {113}},\ \bibinfo {pages} {11324}
  (\bibinfo {year} {2000}{\natexlab{b}})}\BibitemShut {NoStop}%
\bibitem [{\citenamefont {Gallo}\ \emph {et~al.}(2012)\citenamefont {Gallo},
  \citenamefont {Rovere},\ and\ \citenamefont {Chen}}]{Gallo:2012}%
  \BibitemOpen
  \bibfield  {author} {\bibinfo {author} {\bibfnamefont {P.}~\bibnamefont
  {Gallo}}, \bibinfo {author} {\bibfnamefont {M.}~\bibnamefont {Rovere}}, \
  and\ \bibinfo {author} {\bibfnamefont {S.-H.}\ \bibnamefont {Chen}},\ }\href
  {http://stacks.iop.org/0953-8984/24/i=6/a=064109} {\bibfield  {journal}
  {\bibinfo  {journal} {J. Phys.: Condens. Matter}\ }\textbf {\bibinfo {volume}
  {24}},\ \bibinfo {pages} {064109} (\bibinfo {year} {2012})}\BibitemShut
  {NoStop}%
\bibitem [{\citenamefont {Lang}\ \emph {et~al.}(2010)\citenamefont {Lang},
  \citenamefont {Bo\ifmmode~\mbox{\c{t}}\else \c{t}\fi{}an}, \citenamefont
  {Oettel}, \citenamefont {Hajnal}, \citenamefont {Franosch},\ and\
  \citenamefont {Schilling}}]{Lang:2010}%
  \BibitemOpen
  \bibfield  {author} {\bibinfo {author} {\bibfnamefont {S.}~\bibnamefont
  {Lang}}, \bibinfo {author} {\bibfnamefont {V.}~\bibnamefont
  {Bo\ifmmode~\mbox{\c{t}}\else \c{t}\fi{}an}}, \bibinfo {author}
  {\bibfnamefont {M.}~\bibnamefont {Oettel}}, \bibinfo {author} {\bibfnamefont
  {D.}~\bibnamefont {Hajnal}}, \bibinfo {author} {\bibfnamefont
  {T.}~\bibnamefont {Franosch}}, \ and\ \bibinfo {author} {\bibfnamefont
  {R.}~\bibnamefont {Schilling}},\ }\href {\doibase
  10.1103/PhysRevLett.105.125701} {\bibfield  {journal} {\bibinfo  {journal}
  {Phys. Rev. Lett.}\ }\textbf {\bibinfo {volume} {105}},\ \bibinfo {pages}
  {125701} (\bibinfo {year} {2010})}\BibitemShut {NoStop}%
\bibitem [{\citenamefont {G{\"o}tze}\ and\ \citenamefont
  {Sj\"ogren}(1995)}]{Goetze:1995}%
  \BibitemOpen
  \bibfield  {author} {\bibinfo {author} {\bibfnamefont {W.}~\bibnamefont
  {G{\"o}tze}}\ and\ \bibinfo {author} {\bibfnamefont {L.}~\bibnamefont
  {Sj\"ogren}},\ }\href {\doibase 10.1006/jmaa.1995.1352} {\bibfield  {journal}
  {\bibinfo  {journal} {J. Math. Anal. Appl.}\ }\textbf {\bibinfo {volume}
  {195}},\ \bibinfo {pages} {230} (\bibinfo {year} {1995})}\BibitemShut
  {NoStop}%
\bibitem [{\citenamefont {Franosch}\ and\ \citenamefont {$\text{Th.}$
  Voigtmann}(2002)}]{Franosch:2002}%
  \BibitemOpen
  \bibfield  {author} {\bibinfo {author} {\bibfnamefont {T.}~\bibnamefont
  {Franosch}}\ and\ \bibinfo {author} {\bibnamefont {$\text{Th.}$ Voigtmann}},\
  }\href {\doibase 10.1023/A:1019991729106} {\bibfield  {journal} {\bibinfo
  {journal} {{J. Stat. Phys.}}\ }\textbf {\bibinfo {volume} {{109}}},\ \bibinfo
  {pages} {{237}} (\bibinfo {year} {2002})}\BibitemShut {NoStop}%
\bibitem [{\citenamefont {Hansen}\ and\ \citenamefont
  {McDonald}(2006)}]{Hansen:Theory_of_Simple_Liquids}%
  \BibitemOpen
  \bibfield  {author} {\bibinfo {author} {\bibfnamefont {J.~P.}\ \bibnamefont
  {Hansen}}\ and\ \bibinfo {author} {\bibfnamefont {I.~R.}\ \bibnamefont
  {McDonald}},\ }\href@noop {} {\emph {\bibinfo {title} {Theory of Simple
  Liquids}}}\ (\bibinfo  {publisher} {Academic},\ \bibinfo {year}
  {2006})\BibitemShut {NoStop}%
\bibitem [{\citenamefont {Forster}(1975)}]{Forster:Hydrodynamic_Fluctuations}%
  \BibitemOpen
  \bibfield  {author} {\bibinfo {author} {\bibfnamefont {D.}~\bibnamefont
  {Forster}},\ }\href@noop {} {\emph {\bibinfo {title} {Hydrodynamic
  Fluctuations, Broken Symmetry, And Correlation Functions}}}\ (\bibinfo
  {publisher} {Benjamin},\ \bibinfo {year} {1975})\BibitemShut {NoStop}%
\bibitem [{Note1()}]{Note1}%
  \BibitemOpen
  \bibinfo {note} {Since it is clear from the context when $z$ refers to a
  complex frequency or to a distance to the wall, no confusion
  arises.}\BibitemShut {Stop}%
\bibitem [{Note2()}]{Note2}%
  \BibitemOpen
  \bibinfo {note} {In our previous work, Ref.\cite {Lang:2010} we considered
  only symmetric walls, where $[\protect \mathbf {v}]_{\mu \nu } = v_{\nu -\mu
  }= v_{\mu -\nu }$}\BibitemShut {NoStop}%
\bibitem [{\citenamefont {Letz}\ and\ \citenamefont
  {Schilling}(1999)}]{Letz:1999}%
  \BibitemOpen
  \bibfield  {author} {\bibinfo {author} {\bibfnamefont {M.}~\bibnamefont
  {Letz}}\ and\ \bibinfo {author} {\bibfnamefont {R.}~\bibnamefont
  {Schilling}},\ }\href {\doibase 10.1080/13642819908223064} {\bibfield
  {journal} {\bibinfo  {journal} {Phil. Mag. B}\ }\textbf {\bibinfo {volume}
  {79}},\ \bibinfo {pages} {1815} (\bibinfo {year} {1999})}\BibitemShut
  {NoStop}%
\bibitem [{\citenamefont {Theis}\ and\ \citenamefont
  {Schilling}(1998)}]{Theis:1998}%
  \BibitemOpen
  \bibfield  {author} {\bibinfo {author} {\bibfnamefont {C.}~\bibnamefont
  {Theis}}\ and\ \bibinfo {author} {\bibfnamefont {R.}~\bibnamefont
  {Schilling}},\ }\href {\doibase 10.1016/S0022-3093(98)00633-4} {\bibfield
  {journal} {\bibinfo  {journal} {J. Non-Cryst. Solids}\ }\textbf {\bibinfo
  {volume} {235-237}},\ \bibinfo {pages} {106 } (\bibinfo {year}
  {1998})}\BibitemShut {NoStop}%
\bibitem [{\citenamefont {Arnol{\textquoteright}d}(1975)}]{Arnold:1975}%
  \BibitemOpen
  \bibfield  {author} {\bibinfo {author} {\bibfnamefont {V.~I.}\ \bibnamefont
  {Arnol{\textquoteright}d}},\ }\href@noop {} {\bibfield  {journal} {\bibinfo
  {journal} {Russ. Math. Surv.}\ }\textbf {\bibinfo {volume} {30}},\ \bibinfo {pages} {1}
  (\bibinfo {year} {1975})}\BibitemShut {NoStop}%
\bibitem [{\citenamefont {Nyg\aa{}rd}\ \emph {et~al.}(2012)\citenamefont
  {Nyg\aa{}rd}, \citenamefont {Kjellander}, \citenamefont {Sarman},
  \citenamefont {Chodankar}, \citenamefont {Perret}, \citenamefont
  {Buitenhuis},\ and\ \citenamefont {van~der Veen}}]{Nygard:2012}%
  \BibitemOpen
  \bibfield  {author} {\bibinfo {author} {\bibfnamefont {K.}~\bibnamefont
  {Nyg\aa{}rd}}, \bibinfo {author} {\bibfnamefont {R.}~\bibnamefont
  {Kjellander}}, \bibinfo {author} {\bibfnamefont {S.}~\bibnamefont {Sarman}},
  \bibinfo {author} {\bibfnamefont {S.}~\bibnamefont {Chodankar}}, \bibinfo
  {author} {\bibfnamefont {E.}~\bibnamefont {Perret}}, \bibinfo {author}
  {\bibfnamefont {J.}~\bibnamefont {Buitenhuis}}, \ and\ \bibinfo {author}
  {\bibfnamefont {J.~F.}\ \bibnamefont {van~der Veen}},\ }\href {\doibase
  10.1103/PhysRevLett.108.037802} {\bibfield  {journal} {\bibinfo  {journal}
  {Phys. Rev. Lett.}\ }\textbf {\bibinfo {volume} {108}},\ \bibinfo {pages}
  {037802} (\bibinfo {year} {2012})}\BibitemShut {NoStop}%
\bibitem [{\citenamefont {Szamel}\ and\ \citenamefont
  {Flenner}(2010)}]{Szamel:2010}%
  \BibitemOpen
  \bibfield  {author} {\bibinfo {author} {\bibfnamefont {G.}~\bibnamefont
  {Szamel}}\ and\ \bibinfo {author} {\bibfnamefont {E.}~\bibnamefont
  {Flenner}},\ }\href {\doibase 10.1103/PhysRevE.81.031507} {\bibfield
  {journal} {\bibinfo  {journal} {Phys. Rev. E}\ }\textbf {\bibinfo {volume}
  {81}},\ \bibinfo {pages} {031507} (\bibinfo {year} {2010})}\BibitemShut
  {NoStop}%
\bibitem [{\citenamefont {Rajan}\ \emph {et~al.}(1978)\citenamefont {Rajan},
  \citenamefont {Woo},\ and\ \citenamefont {Wu}}]{Rajan:1978}%
  \BibitemOpen
  \bibfield  {author} {\bibinfo {author} {\bibfnamefont {V.~T.}\ \bibnamefont
  {Rajan}}, \bibinfo {author} {\bibfnamefont {C.-W.}\ \bibnamefont {Woo}}, \
  and\ \bibinfo {author} {\bibfnamefont {F.~Y.}\ \bibnamefont {Wu}},\ }\href
  {\doibase 10.1063/1.523719} {\bibfield  {journal} {\bibinfo  {journal} {J.
  Math. Phys.}\ }\textbf {\bibinfo {volume} {19}},\ \bibinfo {pages} {892}
  (\bibinfo {year} {1978})}\BibitemShut {NoStop}%
\bibitem [{\citenamefont {Sciortino}\ and\ \citenamefont
  {Kob}(2001)}]{Sciortino:2001}%
  \BibitemOpen
  \bibfield  {author} {\bibinfo {author} {\bibfnamefont {F.}~\bibnamefont
  {Sciortino}}\ and\ \bibinfo {author} {\bibfnamefont {W.}~\bibnamefont
  {Kob}},\ }\href {\doibase 10.1103/PhysRevLett.86.648} {\bibfield  {journal}
  {\bibinfo  {journal} {Phys. Rev. Lett.}\ }\textbf {\bibinfo {volume} {86}},\
  \bibinfo {pages} {648} (\bibinfo {year} {2001})}\BibitemShut {NoStop}%
\bibitem [{\citenamefont {Szamel}\ and\ \citenamefont
  {L\"owen}(1991)}]{Szamel:1991}%
  \BibitemOpen
  \bibfield  {author} {\bibinfo {author} {\bibfnamefont {G.}~\bibnamefont
  {Szamel}}\ and\ \bibinfo {author} {\bibfnamefont {H.}~\bibnamefont
  {L\"owen}},\ }\href {\doibase 10.1103/PhysRevA.44.8215} {\bibfield  {journal}
  {\bibinfo  {journal} {Phys. Rev. A}\ }\textbf {\bibinfo {volume} {44}},\
  \bibinfo {pages} {8215} (\bibinfo {year} {1991})}\BibitemShut {NoStop}%
\bibitem [{\citenamefont {Franosch}\ \emph {et~al.}(1998)\citenamefont
  {Franosch}, \citenamefont {G\"otze}, \citenamefont {Mayr},\ and\
  \citenamefont {Singh}}]{Franosch:1998a}%
  \BibitemOpen
  \bibfield  {author} {\bibinfo {author} {\bibfnamefont {T.}~\bibnamefont
  {Franosch}}, \bibinfo {author} {\bibfnamefont {W.}~\bibnamefont {G\"otze}},
  \bibinfo {author} {\bibfnamefont {M.}~\bibnamefont {Mayr}}, \ and\ \bibinfo
  {author} {\bibfnamefont {A.}~\bibnamefont {Singh}},\ }\href {\doibase
  10.1016/S0022-3093(98)00500-6} {\bibfield  {journal} {\bibinfo  {journal} {J.
  Non-Cryst. Solids}\ }\textbf {\bibinfo {volume} {235-237}},\ \bibinfo {pages}
  {71 } (\bibinfo {year} {1998})}\BibitemShut {NoStop}%
\bibitem [{\citenamefont {Lang}\ \emph {et~al.}(2012)\citenamefont {Lang},
  \citenamefont {Schilling},\ and\ \citenamefont {Franosch}}]{Lang:2012}%
  \BibitemOpen
  \bibfield  {author} {\bibinfo {author} {\bibfnamefont {S.}~\bibnamefont
  {Lang}}, \bibinfo {author} {\bibfnamefont {R.}~\bibnamefont {Schilling}}, \
  and\ \bibinfo {author} {\bibfnamefont {T.}~\bibnamefont {Franosch}},\
  }\href@noop {} {\bibfield  {journal} {\bibinfo  {journal} {unpublished}\ }
  (\bibinfo {year} {2012})}\BibitemShut {NoStop}%
\bibitem [{\citenamefont {Foffi}\ \emph {et~al.}(2003)\citenamefont {Foffi},
  \citenamefont {G\"otze}, \citenamefont {Sciortino}, \citenamefont
  {Tartaglia},\ and\ \citenamefont {$\text{Th}.$ Voigtmann}}]{Voigtmann:2003}%
  \BibitemOpen
  \bibfield  {author} {\bibinfo {author} {\bibfnamefont {G.}~\bibnamefont
  {Foffi}}, \bibinfo {author} {\bibfnamefont {W.}~\bibnamefont {G\"otze}},
  \bibinfo {author} {\bibfnamefont {F.}~\bibnamefont {Sciortino}}, \bibinfo
  {author} {\bibfnamefont {P.}~\bibnamefont {Tartaglia}}, \ and\ \bibinfo
  {author} {\bibnamefont {$\text{Th}.$ Voigtmann}},\ }\href {\doibase
  10.1103/PhysRevLett.91.085701} {\bibfield  {journal} {\bibinfo  {journal}
  {Phys. Rev. Lett.}\ }\textbf {\bibinfo {volume} {91}},\ \bibinfo {pages}
  {085701} (\bibinfo {year} {2003})}\BibitemShut {NoStop}%
\bibitem [{\citenamefont {$\text{Th}.$ Voigtmann}(2011)}]{Voigtmann:2011}%
  \BibitemOpen
  \bibfield  {author} {\bibinfo {author} {\bibnamefont {$\text{Th}.$
  Voigtmann}},\ }\href {http://stacks.iop.org/0295-5075/96/i=3/a=36006}
  {\bibfield  {journal} {\bibinfo  {journal} {Europhys. Lett.}\ }\textbf
  {\bibinfo {volume} {96}},\ \bibinfo {pages} {36006} (\bibinfo {year}
  {2011})}\BibitemShut {NoStop}%
\bibitem [{\citenamefont {Weysser}\ and\ \citenamefont
  {Hajnal}(2011)}]{Weysser:2011}%
  \BibitemOpen
  \bibfield  {author} {\bibinfo {author} {\bibfnamefont {F.}~\bibnamefont
  {Weysser}}\ and\ \bibinfo {author} {\bibfnamefont {D.}~\bibnamefont
  {Hajnal}},\ }\href {\doibase 10.1103/PhysRevE.83.041503} {\bibfield
  {journal} {\bibinfo  {journal} {Phys. Rev. E}\ }\textbf {\bibinfo {volume}
  {83}},\ \bibinfo {pages} {041503} (\bibinfo {year} {2011})}\BibitemShut
  {NoStop}%
\end{thebibliography}

%

\end{document}